\documentclass[twocolumn]{aastex631}

\shorttitle{Cosmological Fast Radio Burst Dispersion Measures}
\shortauthors{Konietzka et al.}

\usepackage{amsmath}
\usepackage{mathrsfs}
\usepackage{multirow}

\begin{document}

\title{Ray-tracing Fast Radio Bursts Through IllustrisTNG: \\ Cosmological Dispersion Measures from Redshift 0 to 5.5}

\author[0000-0001-8235-2939]{Ralf M. Konietzka}
\altaffiliation{\href{mailto:ralf.konietzka@cfa.harvard.edu}{ralf.konietzka@cfa.harvard.edu}}
\affiliation{Center for Astrophysics $\mid$ Harvard \& Smithsonian, 60 Garden St., Cambridge, MA 02138, USA}

\author[0000-0002-7587-6352]{Liam Connor}
\affiliation{Center for Astrophysics $\mid$ Harvard \& Smithsonian, 60 Garden St., Cambridge, MA 02138, USA}

\author[0000-0002-6648-7136]{Vadim A. Semenov}
\affiliation{Center for Astrophysics $\mid$ Harvard \& Smithsonian, 60 Garden St., Cambridge, MA 02138, USA}

\author[0000-0002-8658-1453]{Angus Beane}
\affiliation{Center for Astrophysics $\mid$ Harvard \& Smithsonian, 60 Garden St., Cambridge, MA 02138, USA}

\author[0000-0001-5976-4599]{Volker Springel}
\affiliation{Max-Planck-Institut für Astrophysik, Karl-Schwarzschild-Str. 1, D-85748, Garching, Germany}

\author[0000-0001-6950-1629]{Lars Hernquist}
\affiliation{Center for Astrophysics $\mid$ Harvard \& Smithsonian, 60 Garden St., Cambridge, MA 02138, USA}

\begin{abstract}

The dispersion measures (DMs) of Fast Radio Bursts (FRBs) arise predominantly from free electrons in the large-scale structure of the Universe. The increasing number of FRB observations have started to empirically constrain the distribution of cosmic baryons, making it crucial to accurately forward model their propagation within cosmological simulations.
In this work, we present a method for measuring FRB DMs in IllustrisTNG that continuously traces rays through the simulation while reconstructing all traversed line segments within the underlying Voronoi mesh.
Leveraging this technique, we create over $20$ publicly available DM catalogs, including a full-sky DM map observed from a Milky Way-like environment.
Our method addresses a problem in previous TNG-based studies, in which a sparse snapshot sampling in the path integral leads to a misestimation of the standard deviation and higher moments of the DM distribution $p(\rm{DM}|z)$ by over $50\%$.
We show that our results are consistent with the most recent observational data from the DSA-110, ASKAP, and CHIME.
We offer a functional form for $p(\rm{DM}| z)$ that provides very good fits across all redshifts from $0$ to $5.5$, showing that the previously proposed log-normal distribution is not well matched to the data.
We find that using simulation box sizes smaller than $35\,h^{-1}\,$Mpc or resolutions with baryonic masses of less than $5\times10^{8}\,h^{-1}\,M_{\odot}$ can distort the derived DM signal by more than $8\%$.
Our findings provide new insights into how the cosmic web shapes FRB signals, and highlight the importance of accurate methodology when comparing cosmological simulations against observations.
\end{abstract}


\section{Introduction} \label{sec:intro}

Fast Radio Bursts (FRBs) are transient bursts in the radio regime at frequencies spanning $0.1$ to $10$\,GHz with a duration of $0.01$ to $10 \,\mathrm{ms}$ \citep{Lorimer_2007, Thornton_2013, Cordes_review2019, Petroff_review2022}.
Since the first detection of an FRB in \citeyear{Lorimer_2007} \citep{Lorimer_2007}, there have been numerous efforts using different telescopes around the world, including CHIME \citep{CHIME_2021}, ASKAP \citep{McConnell_2016_ASKAP}, Parkes \citep{Keith_2010_Parkes}, MeerKAT \citep{Caleb_2023_MeerKAT}, Arecibo \citep{Arecibo_FRB_Spitler_2014}, FAST \citep{Niu_2022_FAST} and the DSA-110 \citep{Law_2024_DSA110}, to increase the number of FRBs.
Today, $\sim\mathcal{O}(10^3)$ FRBs have already been discovered, roughly 100 of which have had sufficient angular \mbox{resolution} to be localized to a host galaxy.

While the association of an FRB at the position of a known magnetar implies that neutron stars with strong magnetic fields are a promising candidate for generating FRBs \citep{Bochenek_2020_magnetar, CHIME1935}, the exact formation channel of these transient events is not fully understood.
In particular, there is an ongoing debate about whether there are other formation channels besides magnetars, possibly evolving older stellar populations, that are capable of producing these radio events \citep{Kirsten_2022_globcluster}.

\begin{figure*}[]
    \centering
    \includegraphics[width=\textwidth]{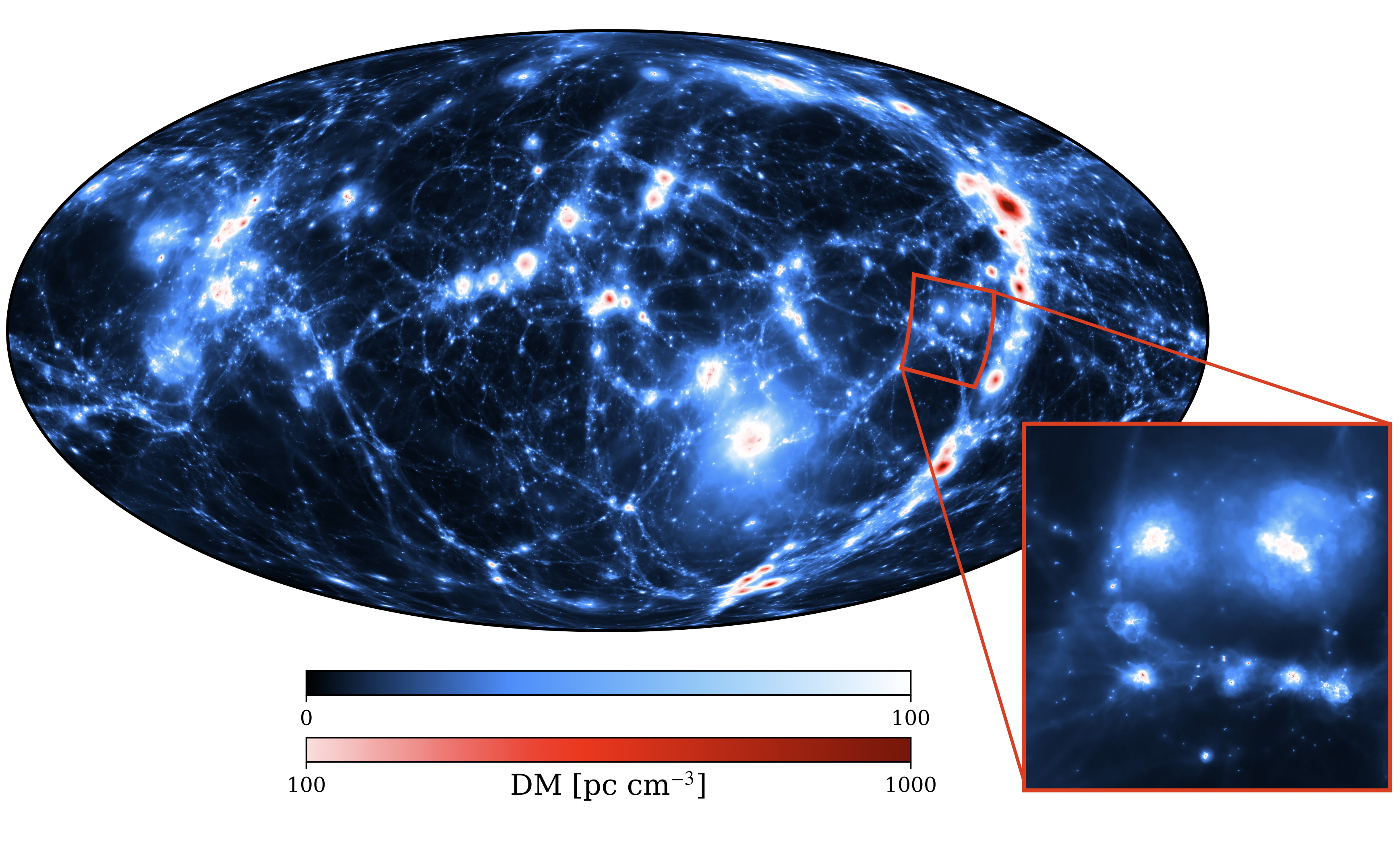}
    \caption{\textbf{Full-sky dispersion measure map integrated up to redshift $\mathbf{0.01}$.}  We show one of our full-sky dispersion measure maps centered on a Milky Way-like environment and integrated up to redshift $0.01$.
    The map was derived from \mbox{TNG300-1} and is described in detail in Section \ref{subsec:continuousmethod} and Section \ref{subsec:DSAsampledata}.
    We see how the three-dimensional structure of the electron density is imprinted in the two-dimensional DM field. Black cosmic voids are traversed by filaments filled with electrons, which are shown in blue, white and red.
    The large foreground galaxy corresponds to an Andromeda analog. The smaller, high-DM clump highlighted with the red color bar resembles a Virgo-like cluster. 
    In the lower right corner, highlighted and framed in red, we zoom into a specific region using our pencil-beam catalog.
    This close-up shows individual halos along one of the filaments and their imprint on the DM signal.
    An evolution of the full-sky map to higher redshifts is shown in Figure \ref{fig:fullsky_evolution} and available at \href{https://ralfkonietzka.github.io/fast-radio-bursts/full-sky-tng/}{https://ralfkonietzka.github.io/fast-radio-bursts/full-sky-tng/}.
    \label{fig:fullskymap}}
\end{figure*}

The characteristic observational feature of all FRBs that makes them a useful probe of the distribution of gas in the Universe is the dispersion of the signal \citep{Cordes_review2019}.
When a radio wave packet travels through a cold plasma with a characteristic plasma frequency $\nu_\mathrm{p}$, longer  wavelengths of the packet take longer to travel through the medium than shorter wavelengths, resulting in a delay of the signal as a function of frequency, called dispersion \citep{Rybicki_Lightman_1979}. 
The strength of the dispersion, called the dispersion measure (DM), is directly proportional to the free electron density $n_\mathrm{e}$ of the plasma, integrated between the source and the observer.
FRBs can therefore be used to probe the gas density between us and the burst origin.
This gas, which is thought to make up the bulk of baryons in the Universe, is otherwise extremely difficult to detect \citep{Fukugita_1998}.

It is then common to decompose the total DM into three contributions, arsing from the free electrons in the Milky Way, the galaxy hosting the FRB, and the intergalactic medium (IGM) in between.
The dispersion measures of local pulsars allow us to constrain the electron distribution in the Milky Way, including the circumgalactic medium and the star-forming Galactic disk \citep{Cordes_2002, Cordes_2003}.
The imprint of the FRB host galaxy can be studied with numerical simulations \citep[see, e.g.,][]{Theis_2024_host} and is expected to be the second largest factor for most sources \citep{Macquart_2020, Connor_2024}.
This work focuses on the largest component: the contribution from the cosmic web, composed of halos, filaments, and voids. 

While up to first order we can approximate the effect of the electrons in the cosmic web using analytical models \citep{Ioka_2003, McQuinn_2014, Fialkov_Loeb_2016}, for higher order deviations we need to take into account the non-linear evolution of the Universe \citep{Hagstotz_2022}.
To tackle this problem we can use large-scale cosmological simulations.
Today, several different large-scale cosmological simulations are available, run with different initial conditions and including different implementations of the underlying physics, such as stellar or black hole feedback \citep[see, e.g.,][for a review]{Vogelsberger_2020}.

To study the effect of the large-scale structure, it is beneficial to use a simulation with a large box size but sufficiently high resolution, capturing a range of small- and large-scale modes. 
In hydrodynamical simulations employing a boxsize of at least $100\,$Mpc, the effect of FRB dispersion measures has previously been analyzed in EAGLE \citep{Batten_2021}, CROCODILE \citep{Zhang_2025_Crocodile}, Illustris \citep{Jaroszynski_2019}, Magneticum \citep{Dolag_2015}, IllustrisTNG \citep{Zhang_2021_DMcatalog,Takahashi_2021,Walker_2024_DMcatalog}, and in simulations by \cite{Zhu_Feng_2021_Ramses} using RAMSES \mbox{\citep{Teyssier_2002}}.

In this work, we use IllustrisTNG \citep{Nelson_2019_TNGrelease}.
We present a method for measuring DMs that continuously traces rays through simulation boxes while reconstructing all of the traversed line segments within the underlying Voronoi mesh.

We compare our method with three other integration techniques to demonstrate that our approach addresses a problem with other TNG-based works \citep{Zhang_2021_DMcatalog, Walker_2024_DMcatalog}.
Previous studies misestimate the standard deviation and higher moments of the DM distribution $p(\mathrm{DM}|z)$ by over $50\%$, due to redshift gaps between simulation snapshots and the method of interpolation between these snapshots.

We study the shape of $p(\mathrm{DM}|z)$ together with the \mbox{DM-redshift} relation, using an FRB DM catalog which is constructed from a plane-parallel cuboid evolving under an inclined angle relative to the simulation box axes.
In this context, we present a new functional form for $p(\mathrm{DM}|z)$ that provides very good fits across all redshifts from 0 to 5.5.

\begin{figure*}[]
    \centering
    \includegraphics[width=1.0\textwidth]{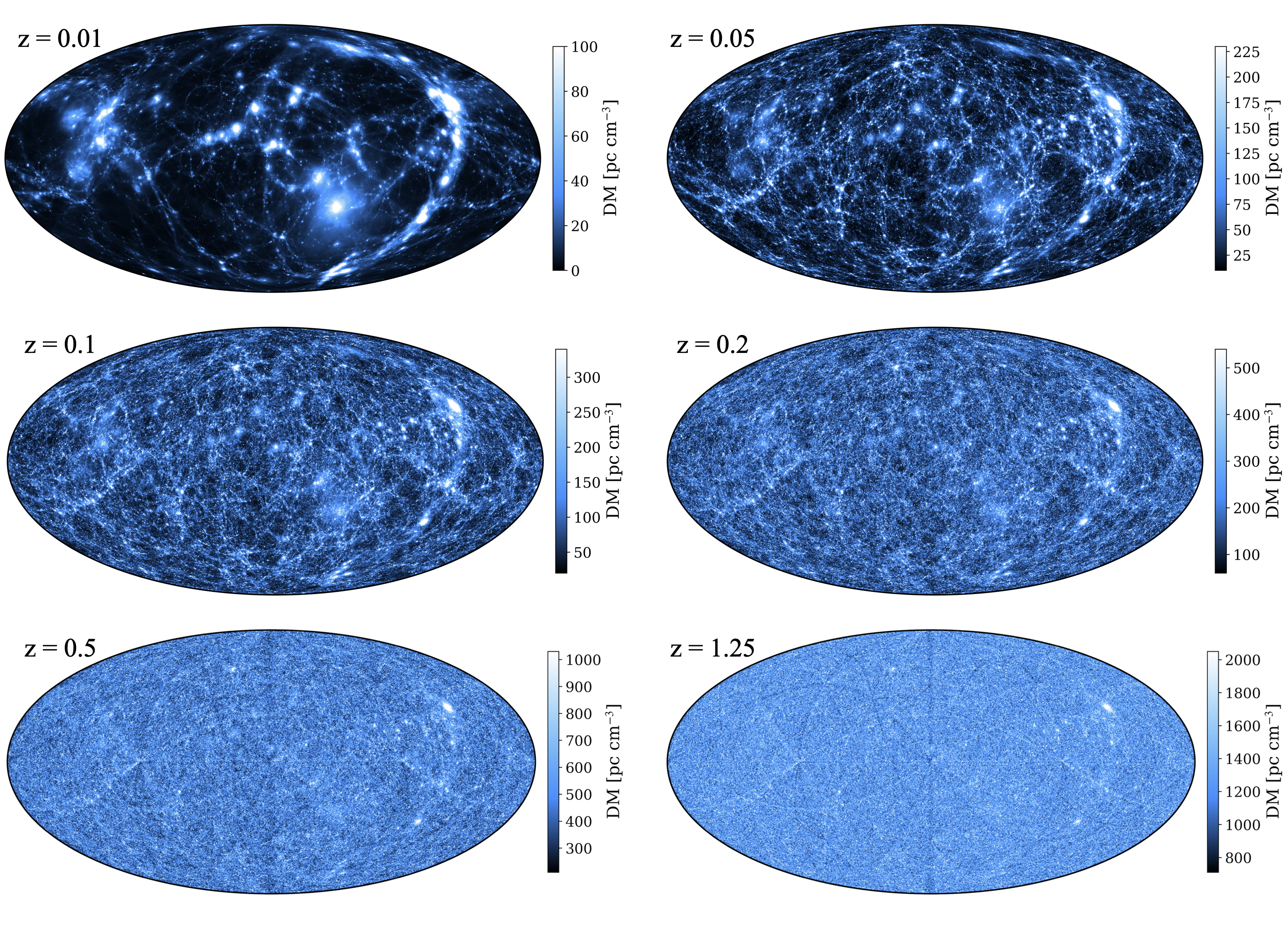}
    \caption{\textbf{Evolution of the full-sky DM map.}
    We show the evolution of our full-sky DM map presented in Figure \ref{fig:fullskymap} with redshift.
    The upper left panel is the same as the map shown in Figure \ref{fig:fullskymap}.
    The more we evolve the map in redshift, the more DM is accumulated. 
    The local environment visible at redshift 0.01 is superimposed by signals from higher redshifts.
    Consequently, the correlations visible at small modes at low redshifts are overlaid by larger modes the farther away we go from the observer’s position.
    At an integration depth of $z=0.5$, we observe some straight lines imprinted on our maps.
    We caution that these are artifacts resulting from the finite simulation volume of TNG (see Section \ref{subsec:choicerayvector} for details).
    A video showing the evolution at smaller steps in redshift can be found at \href{https://ralfkonietzka.github.io/fast-radio-bursts/full-sky-tng/}{https://ralfkonietzka.github.io/fast-radio-bursts/full-sky-tng/}.
    \label{fig:fullsky_evolution}}
\end{figure*}

While previous hydrodynamics-based studies either derive $p(\mathrm{DM}|z)$ from the electron power spectrum \citep{Zhu_Feng_2021_Ramses}, create a light cone \citep{Dolag_2015, Takahashi_2021}, or stack simulation boxes together to form a cuboid that evolves with redshift \citep{Jaroszynski_2019, Batten_2021, Zhang_2021_DMcatalog, Walker_2024_DMcatalog, Zhang_2025_Crocodile}, in this work, we also construct full-sky DM maps centered on Milky Way-like environments.
Our full-sky DM maps, which incorporate the DMs of the FRBs together with the their locations on the sky, are perfectly suited to be compared against the latest observational surveys such as the DSA-110 \citep{Law_2024_DSA110}, CHIME \citep{CHIME_2021} or ASKAP \citep{McConnell_2016_ASKAP}.
Figure \ref{fig:fullskymap} shows one of our full-sky DM maps together with a pencil beam light cone.  We present the evolution of this map as a function of redshift in \mbox{Figure \ref{fig:fullsky_evolution}}.

In addition, we investigate potential biases that could impact our DM catalogs.
We examine how the location of the FRB starting points affects the observed signal.
We also study how the underlying simulation resolution and box size can alter the DM distribution $p(\mathrm{DM}|z)$.

Our work is structured as follows.
In Section \ref{sec:Theoback} we describe the observable under study, the dispersion measure (DM), and the physical basis behind it.
This theoretical background leads us to Section \ref{sec:SimulatingDM}, where we outline how DMs can be measured in numerical simulations.
In Section \ref{sec:Results}, we present and discuss our final results, encapsulated in over $20$ DM catalogs.
We summarize, conclude, and provide an outlook in Section \ref{sec:summary_outlook}.

\section{Theoretical Background} \label{sec:Theoback}
\subsection{The Dispersion Measure} \label{subsec:DM}
When a radio wave of frequency $\nu$ and wavelength $\lambda$ propagates through a cold, unmagnetized plasma with electron density $n_e$, it obeys the dispersion relation $\nu^2 = \nu_p^2 + c^2\,\lambda^{-2}$, where $c$ is the speed of light and $\nu_p(s)$ is the plasma frequency of the medium, which can vary depending on the position $s$ of the observer \citep{Rybicki_Lightman_1979}. 
Correspondingly, the group velocity $v_{g}$ of the radio signal is a function of the frequency $\nu$ and can be expressed as $v_{g}(\nu, s) = c\,(1\,-\,\nu_p^{2}(s)\,\nu^{-2})^{1/2}$, implying that wave packets with higher frequencies propagate faster through the plasma \citep{Rybicki_Lightman_1979}.  As a consequence, the observed radio signal from a source at distance $s_0$ is dispersed with a frequency-dependent time delay $\Delta t$, which can be expressed as
\begin{equation}\label{eq:deltat1}
    \Delta t = \int_0^{s_0} \frac{ds}{\Delta v_{g}(\nu, s)} \approx \int_0^{s_0} ds \, \frac{ \nu_p^{2}(s)}{2\,c} \, \Delta\left(\frac{1}{\nu^2}\right) ,
\end{equation}
where we used in the second step that for radio waves with frequencies $\sim$GHz we are in the regime where $\nu$ is much larger than $\nu_p$ \citep{Fialkov_Loeb_2016}.
Relating the plasma frequency to the electron density $n_e$ via $\nu_p^2(s) = \frac{e^2}{\pi\,m_e}\,n_e(s)$ \citep{Fialkov_Loeb_2016} we arrive at the expression 
\begin{equation}\label{eq:deltat2}
    \Delta t \approx \frac{e^2}{2\pi c\,m_e}\,\Delta\left(\frac{1}{\nu^2}\right)\,\mathrm{DM}(s_0)
\end{equation}
where $e$ is the elementary charge, $m_e$ is the electron mass, and the dispersion measure $\mathrm{DM}(s_0)$ is introduced as the integral of the electron density along the line of sight as
\begin{equation}\label{eq:defDM1}
    \mathrm{DM}(s_0) := \int_0^{s_0} ds \, n_e(s)\,.
\end{equation}

As we motivated earlier, it is convenient to decompose the total observed dispersion measure $\mathrm{DM}$ into three parts \citep{Connor_2024}, the local contribution from the Milky Way's interstellar medium and halo, $\mathrm{DM}_{\mathrm{MW}}$, the contribution from the FRB host galaxy, $\mathrm{DM_{\mathrm{host}}}$, as well as a cosmological component from the ionized medium in the cosmic web, $\mathrm{DM_{\mathrm{cos}}}$, leading to
\begin{equation}\label{eq:defDM_split}
    \mathrm{DM}\left(z\right) = \mathrm{DM}_{\mathrm{MW}}+ \mathrm{DM_{\mathrm{cos}}}\left(z\right) + \mathrm{DM_{\mathrm{host}}}\left(z\right),
\end{equation}
where $z$ is the redshift of the source.
We note that each DM component also depends on its position in the sky.

In addition, in this work, we split $\mathrm{DM_{\mathrm{cos}}}(z)$ into two components.
First, an environment component $\mathrm{DM_{\mathrm{cos,\,env}}}(z)$ that includes the correlation of the FRB emergence position with the comic web.
Second, a statistical, general contribution that is independent of the FRB origin and determined solely by the overall nature of the IGM, called $\mathrm{DM_{\mathrm{cos,\,stat}}}(z)$. This gives: 
\begin{equation}\label{eq:defDM_split_2}
    \mathrm{DM}_{\mathrm{cos}}\left(z\right) = \mathrm{DM_{\mathrm{cos,\,env}}}\left(z\right) + \mathrm{DM_{\mathrm{cos,\,stat}}}\left(z\right).
\end{equation}

While the contribution from the Milky Way is independent of redshift and can be calculated using equation (\ref{eq:defDM1}), for the extragalactic contributions we have to take into account the expansion of the Universe, which stretches the signal with time \citep{Macquart_2020}.

This implies that we have to modify equation (\ref{eq:defDM_split}) with regard to the second and third term, for which the expansion of the Universe cannot be neglected.
Due to the frequency squared dependence in the time delay in equation (\ref{eq:deltat1}), we get an additional factor of $(1+z)^{-1}$ in the integral along the line of sight $s$.
This implies that $\mathrm{DM_{\mathrm{host}}}(z) = \mathrm{DM_{\mathrm{host},0}}\,(1+z)^{-1}$, where $\mathrm{DM_{\mathrm{host},0}}$ is the rest-frame host dispersion measure.
Accordingly, for the cosmological component $\mathrm{DM_{\mathrm{cos}}}(z)$ we obtain
\begin{equation}\label{eq:defDM2}
    \mathrm{DM_{\mathrm{cos}}}(z) = \int_0^z dz' \frac{c\,n_e(z')}{\left(1+z'\right)^2\,H(z')},
\end{equation}
where we used $ds = - dz\,c\,(1+z)^{-1} H^{-1}(z)$ with the Hubble parameter $H(z) = H_0\,E(z)$, which is a function of the Hubble constant $H_0$ and the expansion function $E(z)$.
Note that $n_e$ only includes the electrons that give rise to dispersion. This means that if we want to relate $n_e$ to the total number of baryons in the Universe, we must first account for electrons that are trapped in stars or neutral gas and do not contribute to the dispersion of the signal.

\subsection{Analytic Calculation of the Dispersion Measure}\label{subsec:AnalyticalDM}

In this section we calculate the mean of the integral defined by equation (\ref{eq:defDM2}).
To this end, we have to relate the mean dispersive electron density $\left<n_e\right>(z)$ to the density of all baryonic matter $\rho_b(z)$. We do so by introducing the function $f_{eb}(z)$, the fraction of dispersive electrons to the total number of baryons.
The function $f_{eb}(z)$ encapsulates the strength of the stellar and AGN feedback present in our Universe as well as the evolution of the epoch of reionization.
This leads to:
\begin{equation}\label{eq:fractioneb}
    \left<n_e\right>(z) = f_{eb}(z) \, \frac{\rho_b(z)}{m_p},
\end{equation}
where $m_p$ is the proton mass.
This allows us to express $\left<n_e\right>(z)$ as a function of the fractional baryon density $\Omega_b$ and the Hubble constant $H_0$:
\begin{equation}\label{eq:edensity}
    \left<n_e\right>(z) = \frac{3\,H_0^2\,\Omega_{b}}{8\,\pi\,G\,m_p}\,f_{eb}(z) \, (1+z)^3, 
\end{equation}
where we related $\rho_b(z)$ to $\Omega_{b}$ using the critical density.
$G$ is the gravitational constant.
With an expression of $\left<n_e\right>(z)$ in hand, we rewrite equation (\ref{eq:defDM2}) to highlight the underlying effects that play into the equation:
\begin{equation}\label{eq:defDM3}
    \left<\mathrm{DM_{\mathrm{cos}}}\right>(z) = \int_0^z dz' c_0\,\frac{(1+z')\,\Omega_b \, H_0}{E(z')}\,f_{eb}(z')
\end{equation}
with the constant $c_0$ defined as
\begin{equation}\label{eq:defconstant0}
    c_0 = \frac{3\,c}{8\,\pi\,G\,m_p}.
\end{equation}
From equation (\ref{eq:defDM3}, see also \cite{Ioka_2003}) we observe that the dispersion measure is generally driven by two components.
First, the underlying cosmology, encapsulated by $(1+z')\,\Omega_b \, H_0 \,E(z')^{-1}$.
Second, the evolution of the dispersive electrons with respect to the total baryonic density with time, described by $f_{eb}(z)$.

While we have a good understanding of the former from the measurements of the Planck satellite \citep{Planck_2016}, the study of the second component requires a full understanding of the underlying baryonic feedback model in combination with the reionization history of the Universe.
Moreover, fluctuations in addition to the mean DM evolution, that contain critical insights into the overall spatial distribution of ionized baryons as well as feedback and its interplay with galaxy formation, are not captured by equation (\ref{eq:defDM3}).
Therefore, we must resort to hydrodynamical simulations.

In this work we will use IllustrisTNG \citep{Nelson_2019_TNGrelease}, which we describe below. Following the introduction of IllustrisTNG, we will show how to solve \mbox{equation (\ref{eq:defDM2})} using numerical methods.
The analytical result derived in this section, will be contrasted with the numerical values in Section \ref{sec:Results}.

\section{Simulating Dispersion Measures} \label{sec:SimulatingDM}

The objective of this section is to derive a method for determining the integral that defines the dispersion measure (equation \ref{eq:defDM2}) using numerical simulations.
First, we introduce the suite of simulations used in this work: IllustrisTNG.
Next, we describe four techniques for performing ray-tracing through simulations.
Based on these methods, we discuss the selection of the ray vectors, the starting points of the rays, and the observer's position.
Finally, we show how all of the traversed line segments within the Voronoi grid underlying the IllustrisTNG simulation can be reconstructed and how the electron density is extracted from the simulation.

\subsection{IllustrisTNG} \label{subsec:IllustrisTNG}
In this work, we use \textit{The Next Generation Illustris} simulations (IllustrisTNG), a set of large-scale, cosmological, magneto-hydrodynamical simulations \citep{Nelson_2019_TNGrelease}. IllustrisTNG simulates the evolution of the Universe, including dark matter, cosmic gas, stars, and supermassive black holes, from an initial redshift of $z = 127$ to a redshift of $z = 0$, providing the ideal framework for our analysis \citep{Marinacci_2018_TNG, Naiman_2018_TNG, Nelson_2018_TNG, Pillepich_2018_TNG, Springel_2018_TNG}. 
Based on the Illustris galaxy formation model \citep{Vogelsberger_2013_Illustris}, IllustrisTNG treats magnetic fields by including ideal magneto-hydrodynamics \citep{Pakmor_2011_magnetic,Pakmor_2013_magnetic,Pakmor_2014_magnetic,Springel_2018_TNG}.  In addition, an updated AGN kinetic feedback model and a revised supernova wind model were employed compared to Illustris \citep{Weinberger_2017_TNGmethods, Pillepich_2018_TNGmethods, Marinacci_2018_TNG}.
For more details on the IllustrisTNG galaxy formation model, see \cite{Weinberger_2017_TNGmethods} and \cite{Pillepich_2018_TNGmethods}.

The simulation series consists of three cubic volumes of different sizes, ranging from side lengths $L$ of $205\,h^{-1}\,\text{Mpc}$ (TNG300) to $75\,h^{-1}\,\text{Mpc}$ (TNG100) to $35\,h^{-1}\,\text{Mpc}$ (TNG50), where $h$ is defined by the Hubble constant $H_0$ via $H_0 = h\times100\,\text{km}\,\text{s}^{-1}\,\text{Mpc}^{-1}$ \citep{Nelson_2019_TNGrelease}.
For each of these volumes, three baryonic runs with three different resolutions (high, medium, low) were performed, resulting in a total of nine simulations.
Thereby, the number of particles decreases by a factor of $8$ when switching to the next lower resolution \citep{Nelson_2019_TNGrelease}. Following the IllustrisTNG nomenclature, we refer to the highest resolution as -1, e.g. \mbox{TNG300-1}, while medium and low resolutions are referred to as -2 and -3 respectively \citep{Nelson_2019_TNGrelease}.
The cosmology on which IllustrisTNG is based has been chosen to be consistent with the results of the Planck satellite \citep{Planck_2016, Springel_2018_TNG}. This means $\Omega_{m} = 0.3089$, $\Omega_b = 0.0486$, $\Omega_\Lambda = 0.6911$ and $H_0 = 67.74 \text{km}\,\text{s}^{-1}$ \citep{Springel_2018_TNG}.

Since we are interested in studying the effect of large-scale structures and the cosmic web on the FRB DMs, we focus in this work on the largest TNG volume at its highest resolution with a baryonic mass of about $8\times10^{6}\,h^{-1}\,M_{\odot}$, i.e. TNG300-1.
In addition, we test how a lower resolution (while keeping the volume constant) and a smaller volume (while keeping the resolution constant) affect our results.
For the first case we will use TNG300-2 and TNG300-3.
TNG300-2 and \mbox{TNG300-3} use baryonic masses of about $6\times10^{7}\,h^{-1}\,M_{\odot}$ and $5\times10^{8}\,h^{-1}\,M_{\odot}$, respectively.
For the second case, we will use TNG100-2 and TNG50-3, which with baryonic masses of about $8\times10^{6}\,h^{-1}\,M_{\odot}$ and $4\times10^{6}\,h^{-1}\,M_{\odot}$ have about the same resolution as TNG300-1.

Our focus in this work is the number of free electrons that affect the radio signal.
In this context, we emphasize that IllustrisTNG computes the electron abundance of the low-density gas assuming collisional ionization equilibrium in the presence of a UV background \citep[see][for the methodology]{Katz_1996}.
The UV background, which is modeled to be spatially uniform, is capable of ionizing the baryonic matter, especially hydrogen and helium, hence directly affects the electron density $n_e$ \citep{Faucher-Giguer_2009_UVbackground}.
However, in all IllustrisTNG runs reionization occurs instantaneously, with the UV background being turned on at redshift $z=6$ \citep{Pillepich_2018_TNGmethods, Kannan_2022_Thesan}.
This means that, although IllustrisTNG provides results up to redshift $z = 20$ \citep{Nelson_2019_TNGrelease}, due to the limitations imposed by this first-order reionization model, we limit our focus in this work to redshifts $0$ to $5.5$.
Therefore, we work with the first $13$ complete (i.e., containing all available particle fields) snapshots, which are output at redshifts $0.0$, $0.1$, $0.2$, $0.3$, $0.4$, $0.5$, $0.7$, $1.0$, $1.5$, $2.0$, $3.0$, $4.0$, and $5.0$.

\begin{figure*}[]
    \centering
    \includegraphics[width=\textwidth]{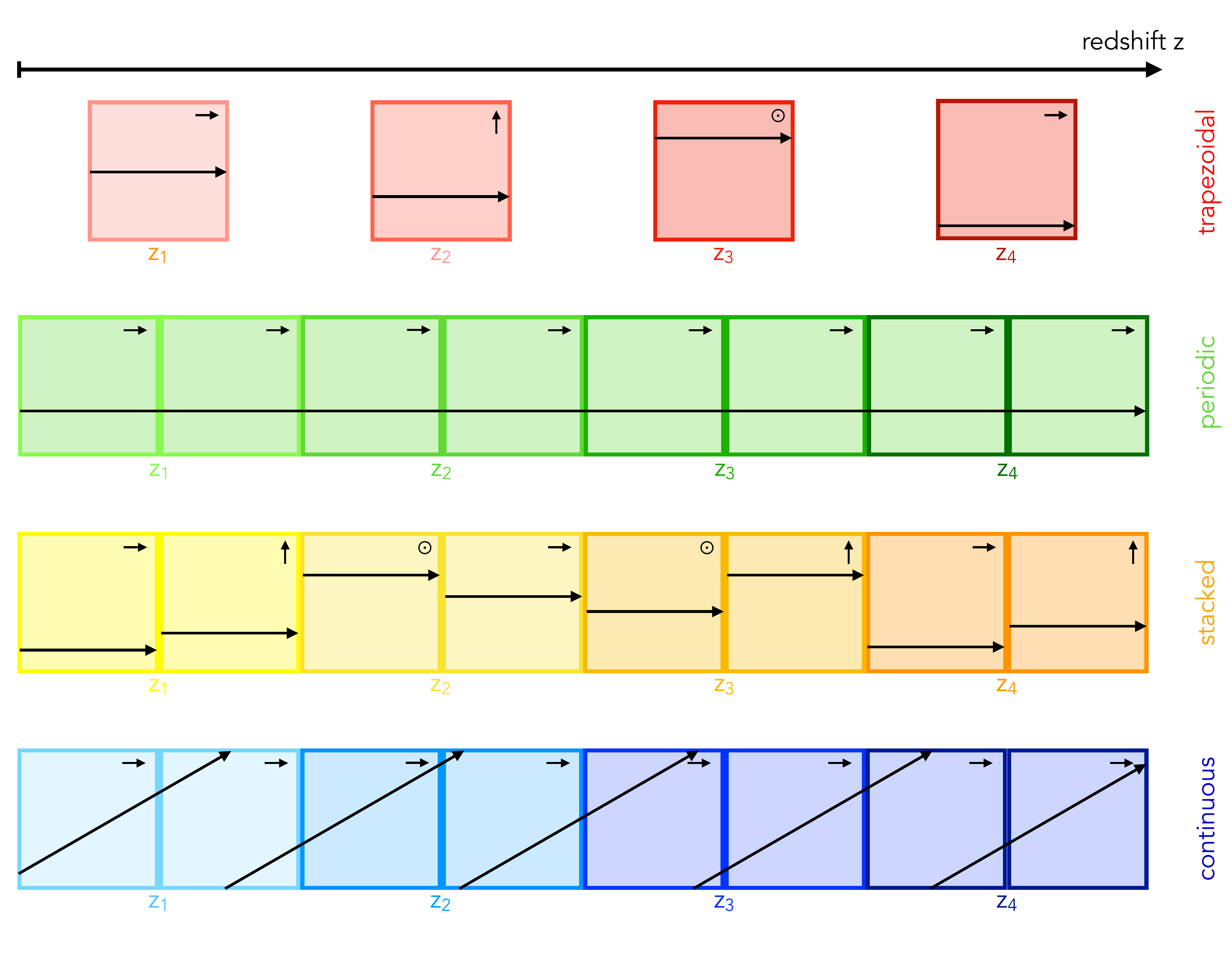}
    \caption{\textbf{Schematic comparison of different methods for calculating dispersion measures in numerical simulations.}
    Redshift increases from left to right.
    Simulation boxes are shown schematically at four different redshifts $z_i$ in four different colors, red, green, orange and blue.
    The large black arrows inside or crossing the boxes represent the directions along which the electron density is integrated.
    The small arrows/symbols in the upper right corner of each box indicate the orientation/rotation of the box in 3D.
    Note that the boxes are assumed to have periodic boundary conditions, so leaving one side of the box will cause the ray to end up on the other side.
    In the top row we show in red the trapezoidal method.
    We see that the boxes are not large enough to fill the entire integration length, resulting in gaps between snapshots.
    These gaps lead to a distortion of the higher moments of the DM signal (see also Sections \ref{subsec:trapeziodalmethod} and \ref{subsec:method_comparison}).
    In the next row we illustrate in green the periodic approach, where all boxes are oriented in the same direction. This implies that the ray will pass through the same features repeatedly, leading to a distortion of the higher moments of the DM signal, as in the case of the trapezoidal method.
    This distortion is avoided by randomly rotating and shifting the simulation boxes, which is displayed in orange.
    We refer to this method as the stacked approach.
    In addition to rotations and shifts, we also randomly shuffle the rays passing through the boxes to avoid correlations between neighboring rays.
    Finally, in the last row, we present the continuous method, which is the ground truth approach used in this work.
    In this case, the ray evolves under an angle through the simulation.
    The tilt of the ray avoids repeating the same structures multiple times, as it occurred in the periodic (green) approach, while allowing the ray to evolve continuously due to the periodic boundary conditions.
    \label{fig:comp_algorithms}}
\end{figure*}

\subsection{Different Ray-tracing Methods}\label{subsec:raytracing}
In this section, we discuss four different methods to solve equation (\ref{eq:defDM2}) using numerical simulations.
We refer to these as the trapezoidal, periodic, stacked and continuous method, respectively.
We schematically compare all four methods in Figure \ref{fig:comp_algorithms} using the colors red (trapezoidal), green (periodic), orange (stacked), and blue (continuous).
We will use this color code throughout our work when referring to a methodology comparison. 

As we outline in Section \ref{subsec:extractingne}, for each snapshot $i$ at redshift $z_i$ (with $i$ running from $1$ to $N$), we can calculate the electron density $n_e(z_i, \vec{x}_{2d})$ of each grid cell at position $\vec{x} = (x_1,x_2,x_3)$ within the simulation.
Hence, we obtain the dispersion measure for a single line of sight and for a single snapshot volume at redshift $z$ using
\begin{equation}\label{eq:DML}
    \mathrm{DM}_{L,\,j}^{(1)}(z, \,\vec{x}_{2d}) = \int_0^L\,\frac{n_{e,\,z}(\vec{x})}{\left(1+z(x_j)\right)^2}\,dx_j,
\end{equation}
where $\vec{x}_{2d} = (x_n, x_m)$ with $n,\,m,\,j$ pairwise different and equal to $1,\,2$ or $3$. $L$ denotes the side length of the simulation box, so $L=205\,h^{-1}\,\text{Mpc}$ for TNG300.

We note that when considering not just a single line of sight at $\vec{x}_{2d}$, but rather several lines of sight, $\mathrm{DM}_{L,j}^{(1)}(z, \vec{x}_{2d})$ changes from a single number to a distribution, represented by $\mathrm{DM}_{L,j}(z)$.
Hereafter, we will drop the subscript $j$, whenever the choice of $j$ does not change or affect our analysis.

Note that all subsequent coordinates or distances refer to comoving coordinates or comoving distances unless stated otherwise.
We denote the commoving distance at redshift $z$ as $\chi(z)$.

\subsubsection{The Trapezoidal Method}\label{subsec:trapeziodalmethod}

The trapezoidal method is based on recent work about FRBs in IllustrisTNG \citep{Zhang_2021_DMcatalog, Walker_2024_DMcatalog}.
Due to the discrete nature of the simulation outputs in redshift, the integral in equation (\ref{eq:defDM2}) needs to be approximated with a discrete sum.
One approach to perform this approximation is the application of the trapezoidal rule.
As a result, \cite{Zhang_2021_DMcatalog} and \cite{Walker_2024_DMcatalog} arrive at the following expression:
\begin{multline}\label{eq:defDM4}
    \mathrm{DM}_{\mathrm{cos}}^{(1)}(z_N) \approx \\
    \approx \sum_{i=1}^{N-1}\left(I(z_{i+1}, \vec{x}_{2d,i+1}) + I(z_{i}, \vec{x}_{2d,i})\right)\,\frac{z_{i+1} - z_i}{2},
\end{multline}
where $I(z_i, \vec{x}_{2d,i})$ is defined as
\begin{equation}\label{eq:integrandDM4}
I(z_i, \vec{x}_{2d,i}) = \frac{c\,\mathrm{DM}_L^{(1)}(z_i, \vec{x}_{2d,i})}{H(z)\,L}.
\end{equation}
We can rewrite equation (\ref{eq:defDM4}) to
\begin{align}\label{eq:defDM5}
    \mathrm{DM}_{\mathrm{cos}}^{(1)}&(z_N) \approx I(z_{1}, \vec{x}_{2d,1})\,\frac{z_{2} - z_{1}}{2} \nonumber\\
    &+ \sum_{i=2}^{N-1}\,I(z_{i}, \vec{x}_{2d,i}) \,\frac{z_{i+1} - z_{i-1}}{2}\nonumber\\
    &+ I(z_N, \vec{x}_{2d,N})\,\frac{z_{N} - z_{N-1}}{2}.
\end{align}

Next, we can consider not just a single line of sight, but an ensemble of lines, implying that we drop the superscript $(1)$ and treat $\mathrm{DM}_{L}(z)$ as a distribution.
Therefore, we have to replace the sums in equation (\ref{eq:defDM5}) with convolutions to arrive at
\begin{equation}\label{eq:defDM6}
    \mathrm{DM_{\mathrm{cos}}}(z_N) \approx \bigotimes_{i=1}^{N}\,g_i\,\mathrm{DM}_L(z_i),
\end{equation}
where we use the symbol $\bigotimes$ as a shorthand for $N$ convolutions and introduced the unitless prefactors $g_i$ as
\begin{multline}\label{eq:prefactorsg}
    g_i = \frac{c}{2\,H(z_i)\,L}\\
    \cdot \left(z_{i+1} - z_{i} + (z_{i} - z_{i-1}) (1-\delta_{i1})(1-\delta_{iN})\right),
\end{multline}
with $\delta_{ij}$ being the Kronecker delta function.
We show schematically this integration method that is based on the trapezoidal rule in Figure \ref{fig:comp_algorithms} in red.

Following the above arguments regarding the trapezoidal rule, we briefly note that one arrives at a slightly different result than that of equation (\ref{eq:defDM6}) if one starts directly from equation (\ref{eq:defDM4}) instead of equation (\ref{eq:defDM5}):
\begin{align}\label{eq:defDM6_alternative}
    \mathrm{DM_{\mathrm{cos}}}&(z_N) \approx \\
    &\approx \bigotimes_{i=1}^{N-1}\left(\tilde{g}_{i+1,1}\mathrm{DM}_L(z_{i+1}) + \tilde{g}_{i0}\mathrm{DM}_L(z_i)\right)
\end{align}
with the unitless prefactors $\tilde{g}_{ij}$ defined as
\begin{equation}\label{eq:prefactorsg_alternative}
    \tilde{g}_{ij} = \frac{c\,\left(\left(z_{i} - z_{i-1}\right)\,\delta_{j1}+\left(z_{i+1} - z_i\right)\,\delta_{j0}\right)}{2\,H(z_i)\,L}.
\end{equation}
In this work we will use equation (\ref{eq:defDM6}) for the trapezoidal approach, as this corresponds to the way the method was implemented previously.

Now, we interpret equation (\ref{eq:defDM6}) and point out the problem that arises with this approach.
Effectively, equation (\ref{eq:defDM6}) represents the linear stacking of $N-1$ boxes of length $L$ into a cuboid of length $(N-1)\,L$, whereby only half of the first and last box is considered. 
However, it is not guaranteed that the comoving length we want to integrate over, i.e. $\chi(z_N)$, is equal to the total length of this cuboid of length $(N-1)\,L$.
In fact, at redshift $z=5.5$, the comoving integration length is $\chi(5.5) \approx 5560\,h^{-1}\,$Mpc, while in TNG300 the length of the cuboid is $(N-1)\,L = 12 \times 205\,h^{-1}\,$Mpc which equals $2460\,h^{-1}\,$Mpc.

To account for this difference in length, the prefactors $g_i$, which are defined such that $\sum_i \,g_i \approx \chi(z_N)\,L^{-1}$, stretch the distribution.
This stretching forces the first moment of the distribution to be correct, but distorts the higher moments of $\mathrm{DM}_L(z)$.
Moreover, even if the two lengths were equal, by the definition of the prefactor $g_i$, the above approach would still distort the distribution because at least $g_0$ and $g_N$ are not equal to $1$.

In other words, this way of accounting for gaps between simulation boxes over-represents outlier lines of sight.
This is especially true when the gaps between snapshots are large or numerous.
Summarizing the above points, we conclude that equation (\ref{eq:defDM6}) does not accurately account for cosmic variance.

To support the previous argument, let us consider a simple example that illustrates the problem.
Let us imagine the Universe as a cuboid of length $L_{U}$.
We simulate two cubic volumes of this Universe with length $L$, where due to the setup of the simulation $M\cdot L = L_{U}$ with $M>0$.
Let us further assume that in each simulated volume $\mathrm{DM}_L$ is normal distributed with some mean $\mu$ and variance $\sigma^2$, implying $\mathrm{DM}_L = \mathscr{N}_L(\mu,\,\sigma^2)$. 
The 'true' $\mathrm{DM_{\mathrm{cos}}}$ distribution of this Universe is then
\begin{equation}\label{eq:DM_true}
    \mathrm{DM_{\mathrm{cos}}} = \bigotimes_{i=1}^{M}\,\mathscr{N}_L(\mu,\,\sigma^2) = \mathscr{N}_L(M\,\mu,\,M\,\sigma^2).
\end{equation}
In contrast, from the approach leading to equation (\ref{eq:defDM6}) we obtain
\begin{align}\label{eq:DM_false}
    \mathrm{DM_{\mathrm{cos}}} &= \left(\frac{M}{2}\mathscr{N}_L\left(\mu,\,\sigma^2\right)\right)
 \ast \left(\frac{M}{2}\mathscr{N}_L\left(\mu,\,\sigma^2\right)\right) \nonumber\\
 &= \mathscr{N}_L\left(M\mu,\,\frac{M^2\,\sigma^2}{2}\right),
\end{align}
where we used that
\begin{equation}\label{eq:normal_stretching}
   \frac{M}{2}\mathscr{N}_L\left(\mu,\,\sigma^2\right) = \mathscr{N}_L\left(\frac{M\,\mu}{2},\,\frac{M^2\,\sigma^2}{4}\right).
\end{equation}
We see that even in this simple case the variance is heavily biased, especially when $M$ becomes large.
As expected, there is no distortion when we simulate the entire Universe, i.e. $M=2$.
This little example motivates the core of this work, which is to continuously ray-trace through the IllustrisTNG simulation to derive full-sky FRB dispersion measure maps that correctly encapsulate higher moments of $\mathrm{DM_{\mathrm{cos}}}$.

As we saw earlier, the side length of the simulation box is usually much smaller than the length of the ray we want to study.
To overcome this problem, one has to fill the gaps between different snapshots.
In general, there are three ways to do this, leading to the three other ray-tracing approaches we describe in the next sections.

\subsubsection{The Periodic Method}\label{subsec:periodicmethod}
For the periodic method (green color in Figure \ref{fig:comp_algorithms}), the rays are defined to be parallel to an axis of the simulation box.
With the periodic boundary conditions of the box, one has constructed an infinitely long ray.
However, with each repetition of a snapshot the ray crosses the same structures repeatedly.
As with the trapezoidal method, this approach will greatly distort the desired distribution.
We can formulate the method as follows.
\begin{equation}\label{eq:defDM7}
    \mathrm{DM_{\mathrm{cos}}}(z_N) \approx \sum_{i=1}^N\,\sum_{k=1}^{K_i}\,\mathrm{DM}_{L_k,\,j}(z_i)
\end{equation}
where we have defined the lengths $L_k$ as well as the integers $K_i$ in such a way that there are no gaps when going from one snapshot to another and $j=1,2$ or $3$ is fixed to a particular axis.
Given that this approach both utilizes and is restricted by the periodicity of the simulation volumes, we refer to the resulting DM catalog as the periodic catalog.

\begin{figure}[h!]
    \centering
    \includegraphics[width=0.45\textwidth]{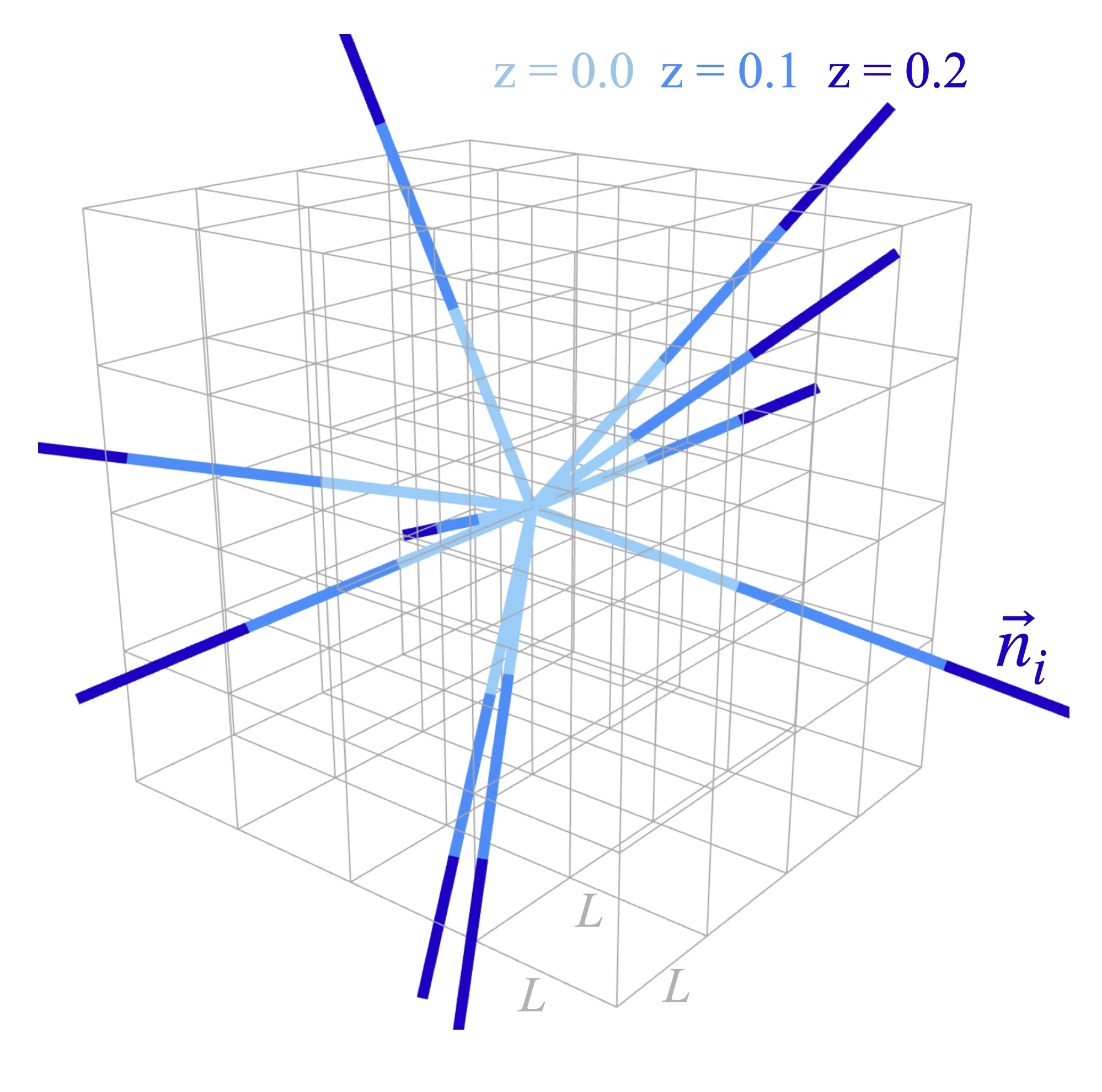}
    \caption{\textbf{Schematic 3D illustration of the continuous method}. We show in grey a grid of simulation boxes. This means that the distance from one neighboring edge to another corresponds to $L$, which is $205\,h^{-1}\,$Mpc for TNG300. In blue we show $10$ random rays going out from the same origin point.
    The color of the rays changes depending on the snapshot file (and corresponding redshift) from which they are drawn.
    A darker blue indicates a higher redshift.
    \label{fig:raytracing_setup}}
\end{figure}

\subsubsection{The Stacked Method}\label{subsec:stackedmethod}
The repetitions apparent in the periodic method can be avoided by randomly rotating and shifting the repeating simulation boxes, and by randomly shuffling the rays passing through them.
In this second way (yellow color in Figure \ref{fig:comp_algorithms}) of filling the gaps between different snapshots, one essentially randomly draws rays from the distribution $\mathrm{DM}_{L,\,j}(z)$ before stacking them, which mathematically translates into
\begin{equation}\label{eq:defDM8}
    \mathrm{DM_{\mathrm{cos}}}(z_N) \approx \bigotimes_{i=1}^{N}\,\bigotimes_{k=1}^{K_i}\,\mathrm{DM}_{L,\,j(k)}(z_i),
\end{equation}
where we introduced random rotations of the boxes by allowing $j(k)$ to vary between $1,2$ and $3$ as a function of $k$.
While this method improves on the first, it comes at the cost of neglecting the periodic boundary conditions of the simulation, which introduces discontinuities that break the correlations within the cosmic web present at the edges of the simulation.
Consequently, we expect this approach to give a good but not ideal result.
This approach is comparatively easy to implement and allows us to compute a large number of rays at low computational cost.
We will refer to the resulting DM catalog as the stacked catalog.

\subsubsection{The Continuous Method}\label{subsec:continuousmethod}
The ideal ray-tracing result is obtained by the following approach.
In this case (blue color in Figure \ref{fig:comp_algorithms}), one avoids the problems of repetition and discontinuities by sending tilted rays through the simulation while using its periodic boundary conditions.
Effectively, this permits a single ray to travel a distance that is much larger ($\gg L$) than the box size $L$ before traversing a similar region twice. 

We implement this approach in three different ways.
First, we construct rays that are all parallel to the same direction vector $\vec{n} = n\,\hat{n}$, starting at different positions in the simulation box.
We discuss the choice of $\vec{n}$ in the next section and will refer to the resulting DM catalog as the continuous catalog.
Second, we create full-sky maps.
This is implemented by having all rays start at the same location in space (the observer's location) and travel outward with different directional vectors $\hat{n}_i$, which are sampled from a HEALPix projection with $N_{\mathrm{side}} = 512$ \citep{Gorski_2005_healpy, Zonca2019_healpy}.
We comment on the choice of the observer's location in Section \ref{subsec:observer_position}.
We will refer to the resulting DM catalog as the full-sky catalog.
Third, we construct a pencil beam to highlight and zoom in on small-scale structures.
We will refer to the resulting DM catalog as the pencil beam catalog.

In all three cases we can write mathematically for a single line
\begin{equation}\label{eq:defDM9}
    \mathrm{DM}_{\mathrm{cos}}^{(1)}(z_N, \vec{\xi}_{2d}) \approx 
    \sum_{i=1}^{N}\,\int_0^{L_i}\,\frac{n_{e\,,z_i}(\vec{\xi})}{\left(1+z_i\right)^2}\,d\xi,
\end{equation}
which translates for an ensemble of lines into
\begin{equation}\label{eq:defDM10}
    \mathrm{DM}_{\mathrm{cos}}(z_N) \approx 
    \sum_{i=1}^{N}\,\int_0^{L_i}\,\frac{n_{e,\,z_i}(\xi)}{\left(1+z_i\right)^2}\,d\xi,
\end{equation}
where the $L_i$ are chosen as 
\begin{equation}\label{eq:length_intboxes}
    L_i = \chi\left(\frac{z_{i+1} - z_i}{2}\right) - L_{i-1},
\end{equation}
with $L_0 = 0$ and $z_1 = 0$, enforcing that there are no gaps between the snapshots.
In Figure \ref{fig:raytracing_setup} we schematically show for $10$ rays how the full-sky continuous catalog was derived.
The lengths of the blue segments correspond to the lengths $L_i$.

\subsection{Choice of the Ray Vector}\label{subsec:choicerayvector}
As \cite{Hilbert_2009_raytracing} pointed out, it is advantageous to choose $\vec{n}=(n_1,\,n_2,\,n_3)$ with co-prime (i.e., with no common divisors greater than $1$) integers $n_i$ and $\vec{n} = n\,\hat{n}$.
This ensures that the plane that is perpendicular to $\hat{n}$ has periodic boundaries encompassing an area of size $n\,L^2$ with simulation box size $L$ \citep{Hilbert_2009_raytracing}.
In addition, the length at which structures will repeat along the line of sight is $L_r = n\,L$ \citep{Hilbert_2009_raytracing}.

\begin{figure}[h!]
    \centering
    \includegraphics[width=0.47\textwidth]{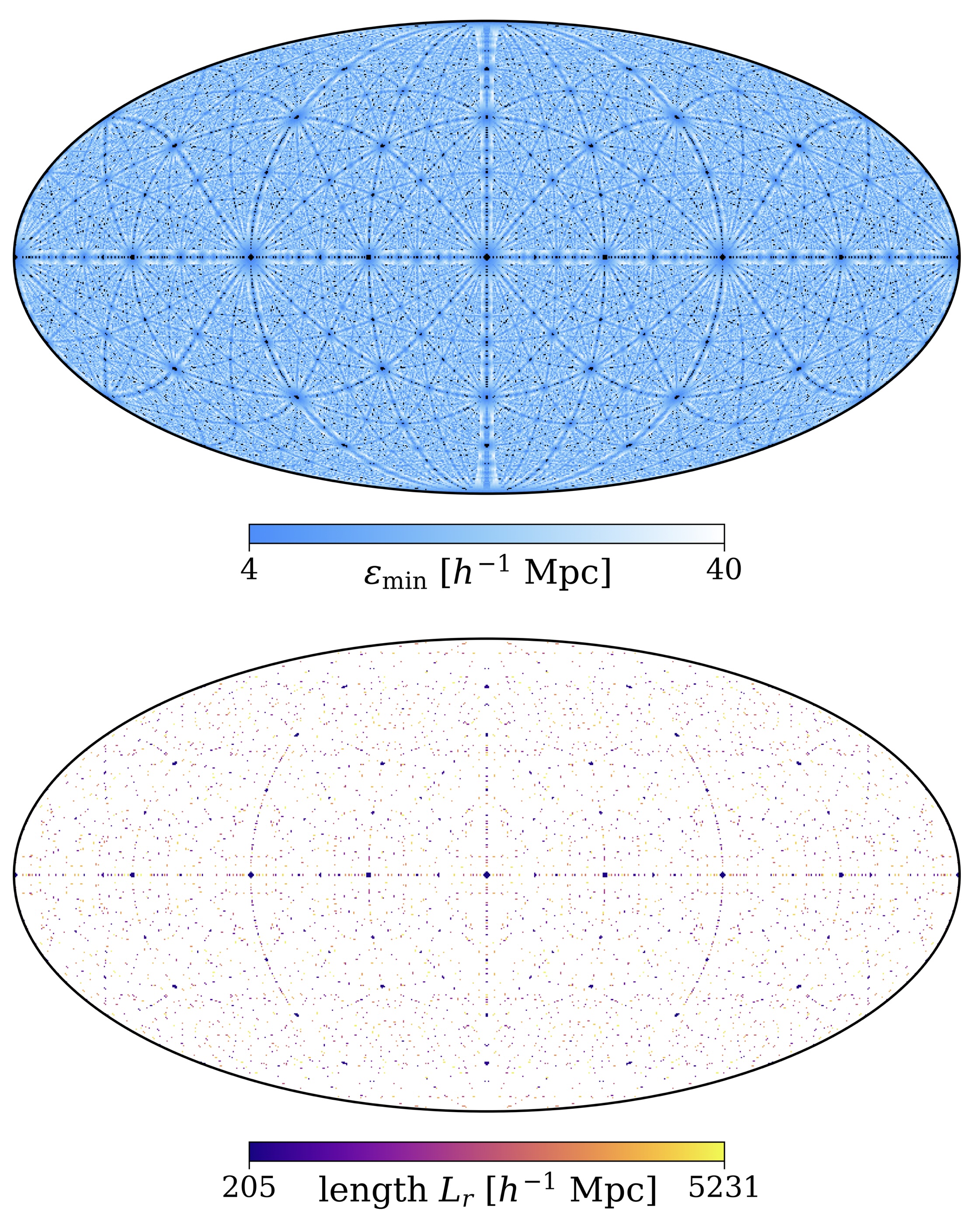}
    \caption{\textbf{The length of repetition and the distribution of $\mathbf{\epsilon_{\mathrm{min}}}$.}
    \textbf{Upper panel.} We show the distance $\epsilon_{\mathrm{min}}$ for a maximum integration length of $5231\,h^{-1}\,$Mpc.
    For $\epsilon_{\mathrm{min}} < 4\,h^{-1}\,$Mpc we mask the image (black dots) to show that no more than $3\%$ of all directions on the sky cross themselves within $4\,h^{-1}\,$Mpc.
    \textbf{Lower panel.} For all the points that were masked in the upper panel as they crossed themselves with $4\,h^{-1}\,$Mpc, we show their length of repetition until this crossing occurs.
    We see that the length at which the intersection takes place is usually greater than $2000\,h^{-1}\,$Mpc.    
    \label{fig:effect_epsilon}}
\end{figure}

The values of the integers $n_i$ depend on two factors.
While we want to maximize the length until we encounter the same position in our grid of simulation boxes a second time, we also want to avoid that the ray comes too close to itself.
The latter refers to the fact that, at any point along the ray, the distance of the ray to itself is smaller than $L$.
This distance may become so small that it is as if the ray had hit itself.

By taking a position $\lambda$ along the ray we can define $\epsilon(\lambda)$ as
\begin{equation}\label{eq:epsilonl}
\epsilon\left(\lambda\right) = \min_{j}\left|\vec{l}_j - \lambda\,\hat{n}\right|,
\end{equation}
where $\vec{l}_j$ denotes a grid point drawn from our grid of simulation boxes.
Given a desired maximum integration length $L_{\mathrm{max}}$, we can find the smallest distance the ray gets to itself, that is $\epsilon_{\mathrm{min}}$, via
\begin{align}\label{eq:epsilonmin}
\epsilon_{\mathrm{min}}\left(L_{\mathrm{max}}\right) &= \min_{\lambda\leqq L_{\mathrm{max}}} \left(\epsilon\left(\lambda\right)\right)\nonumber\\ &= \min_{j,\,\hat{n}\vec{l}_j\leqq L_{\mathrm{max}}}\left|\vec{l}_j - \mathrm{proj}_{\vec{n}}\left(\vec{l}_j\right)\right|,
\end{align}
where $\mathrm{proj_{\cdot}}(\cdot)$ denotes the projection of $\vec{l}_j$ onto $\vec{n}$.

To bring both points together, we want to maximize $L_r$ (ideally $L_{\mathrm{max}} \ll L_r$) while keeping $\epsilon_{\mathrm{min}}(L_{\mathrm{max}})$, the smallest distance between the ray and itself, as large as possible, given a certain integration depth $L_{\mathrm{max}}$.

If we take $n_1 = 1$ and some integer $n_2$, increasing $n_3$ will not change $\epsilon_{\mathrm{min}}$ significantly as long as $n_3$ remains smaller or equal to $n_2^2$.
Once $n_3 > n_2^2$, the distance $\epsilon_{\mathrm{min}}$ will drop to $0$ for increasing $n_3$.
On the other hand, to reach our desired $L_r = \chi(5.5)\approx 5560\,h^{-1}\,$Mpc we need a $|\vec{n}| > 20$.
Therefore, we chose $n_2=5$ and $n_3=25$ to get a long enough $L_r$ with a big enough $\epsilon_{\mathrm{min}}$.
In this case $\epsilon_{\mathrm{min}}$ is about $40\,h^{-1}\,$Mpc.

Furthermore, choosing a single ray direction also implies choosing a single realization of cosmic modes.
We find that this can lead to a change in the distribution of up to $2$-$5\%$.
To avoid this and to correctly model cosmic variance, we therefore use twelve different signed permutations of $\vec{n}$.

The problem of misestimating cosmic variance does not arise with the full-sky catalog as by design this approach effectively covers all possible directions.
However, the problem of the correct choice of the ray direction vector does affect the full-sky catalog.
For some rays, e.g., $\vec{n} = (1,0,0)$, the repetition length $L_{r}$ will become equal to the size of the simulation box $L$.
For these rays, we end up with the same problems as for the periodic catalog.

The question now becomes which fraction of the rays on the full sky are affected and if this effect is large enough to distort the full-sky map.
To tackle this problem we have to redefine the repetition length $L_r$.
We drop the requirement of $n_i$ being co-prime integers, but allow $n_i$ to be any real number. This implies that for some choices of $\hat{n}$, $L_r$ will be going to infinity while $\epsilon_{\mathrm{min}}$ will approach $0$.
Therefore, we should define $L_r$ using a minimum choice of $\epsilon_{min}$:
\begin{equation}\label{eq:repetitionlength}
L_r\left(\epsilon_{min}\right) = \min_{\lambda, \, \epsilon(\lambda)\leqq\epsilon_{min}}\left(\lambda\right).
\end{equation}

According to this definition, a ray crosses itself as soon as it approaches itself within a distance of less than $\epsilon_{\mathrm{min}}$. 
Choosing $\epsilon_{\mathrm{min}} = 4\,h^{-1}\,$Mpc, we see in the upper panel of Figure \ref{fig:effect_epsilon} that for more than $97\%$ of all possible ray directions the rays will reach $5231\,h^{-1}\,$Mpc without intersecting with themselves (using TNG300 with a box size $L = 205\,h^{-1}\,$Mpc).
Even for rays that hit themselves, we see in the lower panel of Figure \ref{fig:effect_epsilon} that the length $L_r$ at which this happens is usually greater than $2000 \,h^{-1}\,$Mpc.
This motivates that ray-tracing in any direction, as done with the full-sky catalog, is consistent with the continuous catalog.
We confirm this in Section \ref{subsec:method_comparison} by comparing the two approaches with each other using TNG300-1.

However, if we use a smaller box with a smaller $L$, the situation changes accordingly.
If we use the TNG50 or TNG100 run, $L$ becomes too small for the continuous approach to avoid repetitions.
In these cases the correlation of the structures with each other will be imprinted in the DM signal as in the periodic catalog.
Therefore, the stacked method is preferred to the continuous method for small box sizes.

\subsection{Starting Points}\label{subsec:startpoints_halo}
In this work, we allow the FRBs to originate from two different ensembles of starting points.
The first ensemble considers FRBs to start from anywhere in the simulation, regardless of the exact initial environment.
This corresponds to $\mathrm{DM_{\mathrm{cos,\,stat}}}(z)$ as described in Section \ref{subsec:DM}.

The second ensemble incorporates the FRB environment.
While we explicitly do not model $\mathrm{DM}_\mathrm{host}$ and $\mathrm{DM}_\mathrm{MW}$ in this work, we are interested in correlations of the extragalactic DM with its host environment.
By this we mean correlations that occur outside the host halo of the FRB.
For example, if the host is a very massive galaxy, there may be an increased probability of crossing a dense filament immediately after the FRB leaves the host.
This corresponds to \mbox{$\mathrm{DM_{\mathrm{cos}}}(z) = \mathrm{DM_{\mathrm{cos,\,stat}}}(z) + \mathrm{DM_{\mathrm{cos,\,env}}}(z)$}.
Therefore, the imprint of the environment $\mathrm{DM_{\mathrm{cos,\,env}}}(z)$ is given as the difference between the two ensembles $\mathrm{DM_{\mathrm{cos}}}(z)$ and $\mathrm{DM_{\mathrm{cos,\,stat}}}(z)$.

\begin{figure}[h!]
    \centering
    \includegraphics[width=0.37\textwidth]{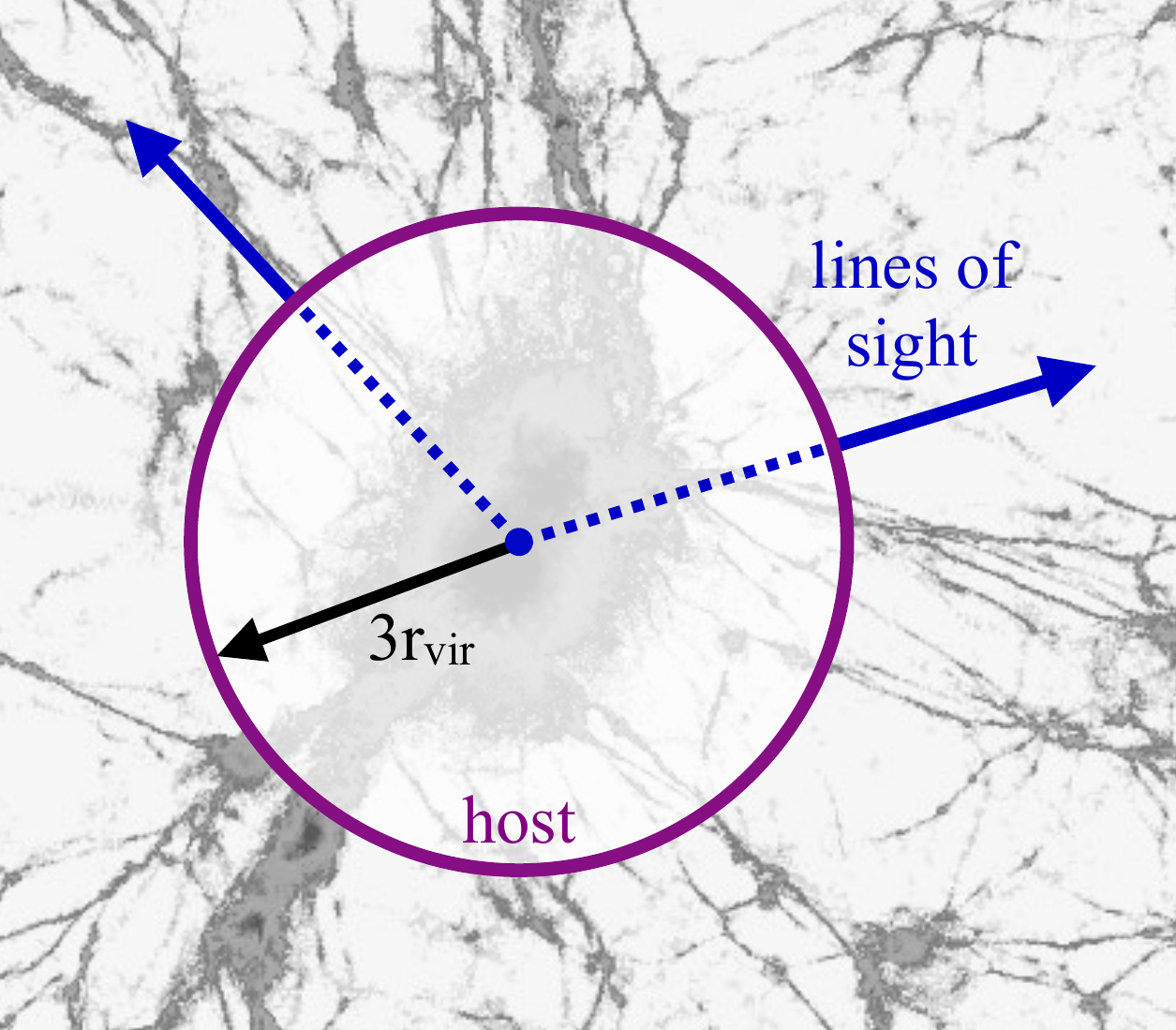}
    \caption{\textbf{Definition of starting points.}  We show how we define the starting points of FRBs when we want to include imprints of the FRB environment, as described in Section \ref{subsec:startpoints_halo}.
    The effect of the FRB source environment corresponds to $\mathrm{DM_{\mathrm{cos}}}(z) - \mathrm{DM_{\mathrm{cos,\,stat}}}(z)$ (see Section \ref{subsec:DM} for details).
    This is in addition to the host contribution.
    We do not constrain or model the host contribution in this work.
    Once we have identified a halo, we position an FRB on its center, represented by the blue point.
    The radio signal can travel in any direction on the 3D sphere, illustrated by the blue arrows. 
    We exclude signal that occurs within the host, indicated by the dashed blue line.
    The host is defined as the region within three virial radii from the halo center.
    The background map shows the gas density distribution for a thin slice through IllustrisTNG \citep{Nelson_2019_TNGrelease}.
    \label{fig:ray_selection}}
\end{figure}

Figure \ref{fig:ray_selection} shows how we implement this.
Once we have identified a halo, we position the FRB at the halo center.
The radio signal can travel in any direction on the 3D sphere.
We exclude signal that occurs within the host, defined by the region within three virial radii from the halo center.
We define the virial radius as the radius of a sphere around the center of a halo whose mean density is equal to $200$ times the critical density of the Universe \citep{Nelson_2019_TNGrelease}.

\begin{figure*}[]
    \centering
    \includegraphics[width=1.0\textwidth]{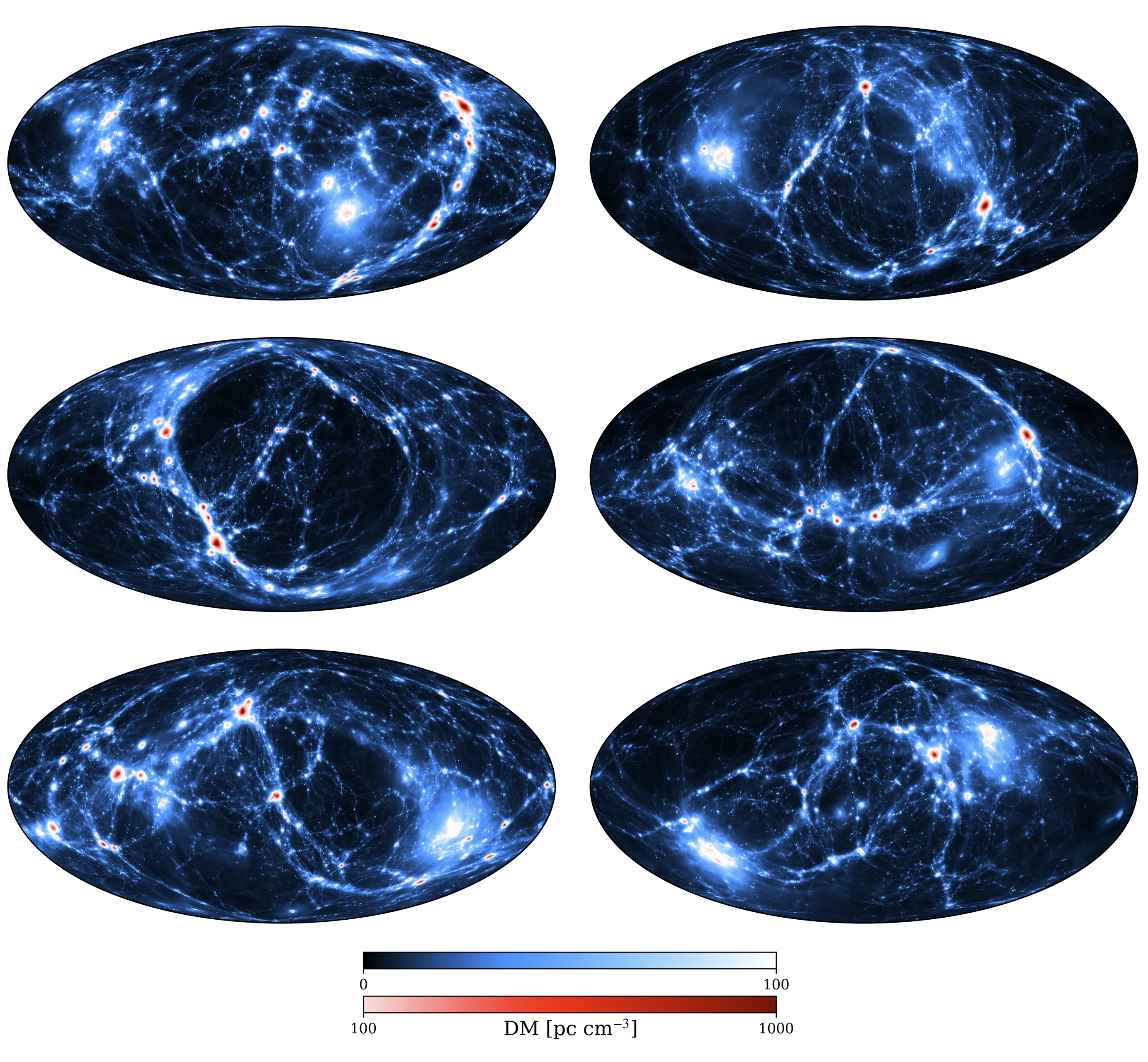}
    \caption{\textbf{Different Milky Way-like environments.}
    We present six different Milky Way-like DM environments integrated up to a redshift of $0.01$.
    The maps are derived from TNG300-1.
    The selection of these galaxies/environments is described in Section \ref{subsec:observer_position}.
    In all six panels, the large foreground galaxy corresponds to an Andromeda analog.
    The smaller, high-DM clump highlighted with the red color bar resembles a Virgo-like cluster.
    The environments in the two uppermost panels were chosen for the deep, full-sky maps that evolve from redshift $0$ to $1.25$.
    The uppermost left panel is the same map as the one shown in Figures \ref{fig:fullskymap} and \ref{fig:fullsky_evolution}.
    All images are created using the continuous ray-tracing method.
    \label{fig:fullsky_observer_change}}
\end{figure*}

We also distinguish between two types of halos: star-forming and massive halos.
To select these halos, we create probability densities based on the star formation rate and stellar mass of all halos in IllustrisTNG.
Using these PDFs, we either choose halos that are more likely to form stars, or halos that are more likely to have a higher stellar mass.

We do all of the above steps at redshifts $0.1$, $0.5$, $1.0$, and $1.5$.
This implies that we create eight different catalogs in total.
\vfill

\subsection{The Observer's Location}\label{subsec:observer_position}
For the full-sky and pencil beam catalogs, we have to choose an observer's location.
To closely mimic observational data, we decided to position the observer within Milky Way-like environments in TNG300.

Our selection of these environments is based on previous studies of Milky Way-like galaxies in IllustrisTNG \citep{Pillepich_2024_MW_analog, Semenov_2024_MW_analog} and can be summarized as follows.

We only consider central (i.e., not satellite) galaxies at a redshift of zero.
These galaxies must have a total stellar mass between $3$ and $8\times10^{10}\,M_{\odot}$ and be embedded in a dark matter halo with a mass between $0.8$ and $1.4\times10^{12}\,M_{\odot}$.
We use the mass enclosed in a spherical overdensity of $200$ times the critical density of the Universe for the mass of the dark matter halo \citep{Nelson_2018_TNG}.
We also require that the central galaxy has Virgo cluster- and Andromeda-like companions, which we describe below.

To mimic the appearance of a Virgo-like cluster \citep{Fouque_2001_Virgo,Mei_2007_Virgo}, we require that exactly one halo with a dark matter mass of at least $10^{14}\,M_{\odot}$ be located between $15$ and $18$ Mpc from the center of the selected galaxy.
We exclude galaxies that have another massive halo with a dark matter mass of at least $10^{13} M_{\odot}$ closer than the Virgo-like halo.

Additionally, to imitate an Andromeda-like galaxy, we require that  no galaxy rich in stellar matter (i.e., with a total stellar mass greater than $3\times10^{10}M_{\odot}$) be within $500$ kpc of the central galaxy, and that at least one galaxy rich in stellar matter be between $500$ and $2500$ kpc from the central galaxy.

These cuts leave us with six galaxies located in Milky Way-like environments within TNG300.
The Andromeda analogues of these galaxies are between $900$ and $2000$ kpc from the center of their respective galaxies.
The analogs of the Virgo cluster are located between $16.7$ and $17.8$ Mpc from the center of the galaxy.
We present all six environments in Figure \ref{fig:fullsky_observer_change}, integrated up to redshift $0.01$.

As in Section \ref{subsec:startpoints_halo}, we exclude signal arising within three virial radii of the center of the selected galaxy, as in this work we focus on modeling the extragalactic DM contribution.
We do not intend to simulate the Milky Way's circumgalactic medium or star-forming Galactic disk \citep{Foley_2023, Konietzka_2024, MiretRoig_2024, Hacar_2025}.

\subsection{Reconstruction of All Traversed Line Segments within the Voronoi Tessellation}\label{subsec:Voronoi}
Here, we describe how to perform the integral in equation (\ref{eq:DML}).
Regardless of the simulation being examined, the integral in equation (\ref{eq:DML}) can always be approximated by a sum over the grid defined by $\Delta\xi_l$ of the simulation along the direction of $\xi_j$.
\begin{equation}\label{eq:intne}
    \int\,n_e(z,\vec{\xi})\,d\xi_j = \sum_{l}\,n_e(z,\vec{\xi}_{2d},\xi_{j,k})\,\Delta\xi_l .
\end{equation}
Note that, compared to equation (\ref{eq:DML}), in the equation above we dropped the $(1+z)^{-2}$ term, as it is irrelevant to the discussion below.

\begin{figure}[h]
    \raggedright
    \includegraphics[width=0.47\textwidth]{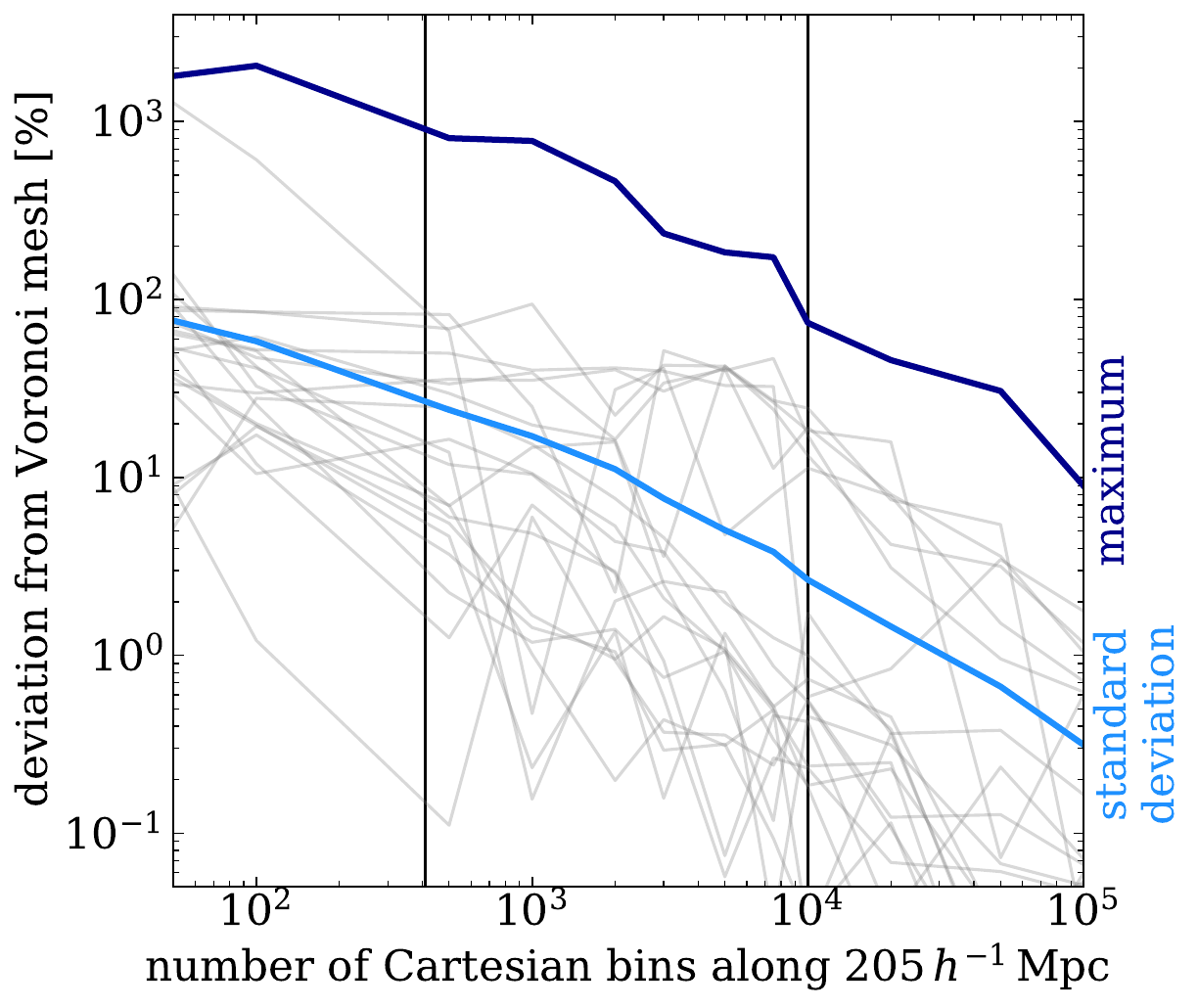}
    \caption{\textbf{Effect of Cartesian gridding.} We show the difference between reconstructing all traversed line segments within the Voronoi tessellation and mapping the Voronoi tessellation onto a Cartesian grid with a certain number of bins.
    We show the number of bins along a ray with a length of $205\,h^{-1}\,$Mpc as the x-axis.  The error introduced by the Cartesian gridding is shown as the y-axis. 
    As expected, we see that the deviation between the two approaches decreases as we increase the number of Cartesian bins.
    With the vertical black lines we show the numbers of bins that were taken in previous studies.
    \cite{Zhang_2021_DMcatalog} and \cite{Walker_2024_DMcatalog} take $10^4$ bins.
    \cite{Cheng_2025} uses $410$ bins.
    The standard deviation between the two approaches is displayed in light blue.
    The maximum induced error is presented in dark blue.
    We see that gridding with $10^4$ bins introduces a bias of about $1$ to up to $10\%$ compared to reconstructing all traversed line segments within the Voronoi mesh.
    Using fewer than $10^3$ bins can bias the results significantly.
    \label{fig:cartesian_gridding}}
\end{figure}

IllustrisTNG was performed with the moving mesh code \textsc{arepo} \citep{Springel_2010_Arepo}. \textsc{arepo} is characterized by the Voronoi tessellation of a set of discrete mesh generation points that span an unstructured moving mesh on which the magnetohydrodynamic equations are solved \citep{Springel_2010_Arepo}.
This non-uniform structure of the Voronoi mesh makes it difficult to analyze the simulation data, as this means that the $\Delta\xi_l$ are different for different values of $l$.

\begin{figure}[]
    \centering
    \includegraphics[width=0.45\textwidth]{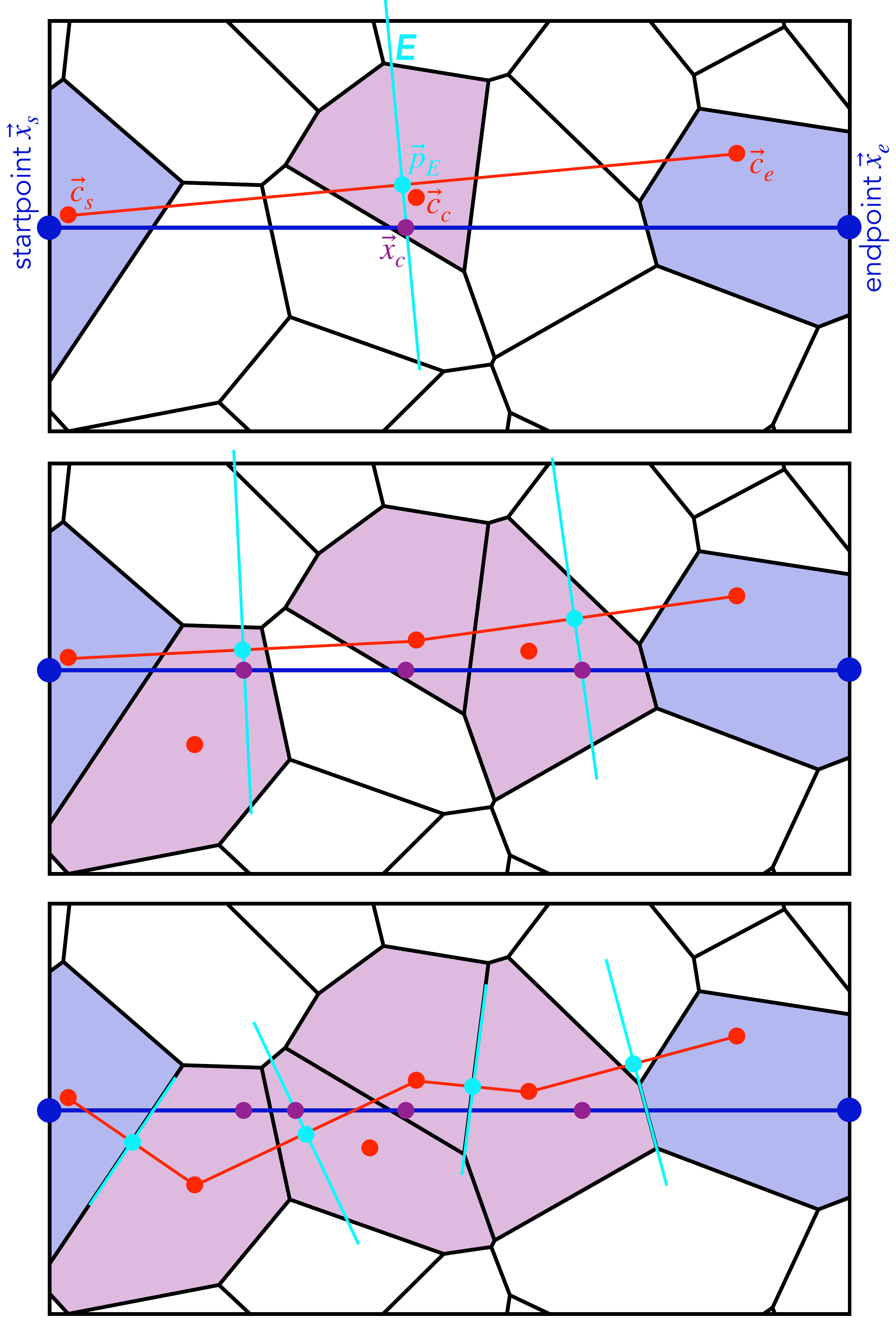}
    \caption{\textbf{Reconstruction of all traversed line segments within the Voronoi tessellation.} We illustrate how we reconstruct all traversed line segments within the Voronoi tessellation.
    \textbf{Upper panel.}
    We show the ray's starting point $\vec{x_s}$ and end point $\vec{x_e}$ in blue.
    The ray develops along the blue line described by the normalized vector $\hat{n}$.
    For both points, start and end, we identify the cell center of the Voronoi cell in which they are located, shown in red and labeled $\vec{c_s}$ and $\vec{c_e}$.
    The points $\vec{c_s}$ and $\vec{c_e}$ define a second ray, colored red, which evolves along the normal vector $\vec{m} = (\vec{c_e} - \vec{c_s})$.
    In light blue, we show the plane $E$ that is defined by the vector $\hat{m}$ and the reference point $\vec{p}_E = \frac{1}{2}(\vec{c_e} + \vec{c_s})$.
    This plane divides the red ray exactly in half.
    Using the definition of $E$, we can calculate the crossing point $\vec{x_c}$ (purple point) of $E$ with the blue ray.
    With $\vec{x_c}$ we have found a new cell and its center $\vec{c_c}$, shown in red, with which we can repeat the above procedure.
    \textbf{Middle panel.}
    Using the center of the new cell we can define two new rays, colored in red.
    We show two additional planes shown in light blue, that cross the red rays at their middle.
    As in the upper panel, we calculate the crossing points of the planes with the blue ray and display them in purple.
    We have found two new cells.
    \textbf{Bottom panel.}
    We repeat the above procedure until we have found all crossed cells crossed by the blue ray.
    Note that in three cases the light blue plane is the boundary plane between two cells.
    This correspondence follows from the definition of the Voronoi mesh.
    For these cells we reached the stopping condition of our algorithm.
    \label{fig:arepo}}
\end{figure}

There are two ways to work with the Voronoi tessellation.
One way is to map the Voronoi tessellation onto a Cartesian grid, implying that $\Delta\xi_{l_1} = \Delta\xi_{l_2}$ for $l_1 \neq l_2$, which simplifies the above sum.
However, this has the disadvantage of losing the information of the densest parts of the simulation if the grid size $\Delta\xi$ is chosen larger than the smallest cell size, which is less than $1$ pc. 
We show in the Figure \ref{fig:cartesian_gridding} the error that this mapping to a Cartesian grid introduces as a function of number of grid bins.
We observe that the selection of Cartesian grids in previous studies introduces biases ranging from about $1\%$ to over $20\%$ \citep{Zhang_2021_DMcatalog, Walker_2024_DMcatalog, Cheng_2025}.

To avoid this bias, in this work, we reproduce all traversed line segments within the underlying Voronoi tessellation.
This ensures that all information is preserved, even in the densest parts of the simulation.

Due to the large snapshot size, prior to reconstructing all traversed line segments within the Voronoi tessellation, one usually preselects all gas cells that are likely to be crossed by a single ray.
This preselection can be done by considering only those gas cells that are within a radius $r_{max}$ of the ray.
Following \cite{Smith_2022_Thesan_rttube}, an appropriate way to select $r_{max}$ is to approximate it by the size of the largest grid cell.
If $V_{max}$ refers to the volume of that largest cell, then a sufficient choice of $r_{max}$ is $r_{max} = \eta\,\sqrt[3]{V_{max}}$ with the factor $\eta$ taking into account the geometry of the cell \citep{Smith_2022_Thesan_rttube}.
For a perfectly spherical cell $\eta = \frac{3}{4\,\pi} \approx 0.62$.
Based on convergence tests, \cite{Smith_2022_Thesan_rttube} employs $\eta = 0.75$.
In this work, we adopt $\eta=1$, resulting in $r\approx0.4\,h^{-1}\,$Mpc.
We note that this value of $r_{max}$ is four times larger than recent choices in the literature. In fact, by choosing $r_{max} = 0.1\,h^{-1}\,$Mpc, one incorrectly excludes up to $10\%$ of the cells in TNG300-1 that the ray actually traverses.

To reduce the amount of data we have to store on the fly, in addition to preselecting the data using a selection of $r_{max}$ based on the largest cell size, we make another selection based on the individual cell volumes.
For a single ray, we include all cells $q$ with volume $V_q$ that have a distance to that ray of less than $r_q = 2\,\sqrt[3]{V_q}$.

With these preselected tubes of cells we can now reconstruct all traversed line segments within the Voronoi tessellation, which we will do recursively.
Let us assume that we have a ray with length $d$ and starting point $\vec{x_s}$, that develops in direction $\hat{n}$.
The complete ray $\vec{r}$ can be described by $\vec{r} = \vec{x_s} + \lambda\,\hat{n}$ where $d \geq \lambda \geq 0$, which implies an end point $\vec{x_e} = d\,\hat{n}$.
For both points, start and end, we identify the mesh generation point or cell center of the Voronoi cell in which they are located.
We denote these centers as $\vec{c_s}$ and $\vec{c_e}$.
The coordinates of the mesh generation points or the positions of the gas cells are provided in IllustrisTNG by the field \textit{Coordinates}.
The points $\vec{c_s}$ and $\vec{c_e}$ define another ray, evolving along the vector $\vec{m} = (\vec{c_e} - \vec{c_s}) = m\,\hat{m}$.
Next, we define a plane $E$ through the vector $\hat{m}$ and the reference point $\vec{p}_E = \frac{1}{2}(\vec{c_e} + \vec{c_s})$.
This plane divides the ray along $\hat{m}$ exactly in half.
Therefore, following the definition of the Voronoi mesh, if the start and end cells were connected, the plane $E$ would be the boundary between the two cells.
Using the definition of $E$, we can calculate the crossing point $\vec{x_c}$ of $E$ with the ray $\vec{r}$ via $\vec{x}_c = \vec{c_s} + \lambda_c\,\hat{n}$ where $\lambda_c$ is defined as
\begin{equation}\label{eq:lambda}
    \lambda_c = \left(\frac{m}{2} - (\vec{c}_s-\vec{x}_s)\,\hat{m}\right)\,\frac{1}{\hat{n}\,\hat{m}}.
\end{equation}
Now, we can check if $\vec{x}_c$ is located inside the cells specified by $\vec{c_s}$ or $\vec{c_e}$.
If this is the case, by construction the plane $E$ is the boundary plane between those cells and we can conclude that there is no additional cell in between those cells.
If not, we have found a new point $\vec{x_c}$ and its cell center $\vec{c_c}$, with which we can repeat the above procedure.
We stop the algorithm once all boundary planes between all cells that were crossed have been identified, such that we know that there is no additional cell in between our start- and endpoint.
We illustrate this recursive algorithm in Figure \ref{fig:arepo}.
A similar algorithm that reconstructs all traversed line segments within a Voronoi mesh is implemented in \textsc{vortrace} (A. Beane $\&$ M. C. Smith, in preparation).

\subsection{Extracting the Electron Density}\label{subsec:extractingne}
There is only one part remaining to calculate the integral defining the dispersion measure given by equation (\ref{eq:DML}): we have to determine the electron density $n_e$.
For low-density gas, below the star formation threshold, IllustrisTNG computes the electron abundance assuming collisional ionization equilibrium \citep{Pillepich_2018_TNGmethods, Katz_1996}.
At higher densities, a subgrid model based on an effective equation of state is used instead \citep{Springel_Hernquist_2003}.
Therefore, we must distinguish between the two cases: Voronoi cells that are star-forming and those that are not, determined by the field \textit{StarFormationRate} in IllustrisTNG.

While for non-star-forming cells we can use the outputs of the simulation, for star-forming cells we have to invert the subgrid model \citep{Springel_Hernquist_2003} in order to derive from it the number of free electrons in that cell.
We perform this inversion and find that including star-forming cells leads to an increase in DM of at most $2\,$pc$\,$cm$^{-3}$ by redshift $5.5$, corresponding to a sub-percent change.
Therefore, setting the ionization fraction of these cells equal to zero has no effect on our conclusions.

For all the results presented in this work, we set the free electron density of star-forming gas to zero and calculate the free electron density of non-star-forming gas as follows
\begin{equation}
n_e = f_{eH}\,\,\frac{x_H\,\,\rho_{gas}}{m_p} \,\,(1+z)^3,
\end{equation}
where $x_H = 0.76$ is the hydrogen fraction, $m_p$ the proton mass, $\rho_{gas}$ the gas mass density and $f_{eH}$ the fractional electron number density with respect to the total hydrogen number density.  The last two quantities are provided in IllustrisTNG by the fields \textit{Density} and \textit{ElectronAbundance}, respectively.

\begin{deluxetable*}{ccccccccc}
\tablecaption{Dispersion Measure Catalogs\label{tab:DMcatalogs}}
\tablehead{
  \colhead{} & \colhead{Name} & \colhead{Method} & \colhead{Ray Direction $\vec{n}$} & 
  \colhead{Start} & \colhead{Observer} & \colhead{Ray Number} & \colhead{Redshift} & \colhead{Simulation}}
\startdata
1 & continuous & continuous & $(1,5,25)$ & random & plane $\perp \vec{n}$ & $1.2\times10^5$ & $5.5$ & TNG300-1 \\
2 & random & continuous & uniform & random & random & $1\times10^5$ & $5.5$ & TNG300-1 \\
3.1 & full-sky 1 & continuous & HEALPix & random & MWE 1 & $3\times10^6$ & $1.25$ & TNG300-1 \\
3.2 & full-sky 2 & continuous & HEALPix & random & MWE 2 & $3\times10^6$ & $1.25$ & TNG300-1 \\
3.3 & full-sky 3 & continuous & HEALPix & random & MWE 3 & $3\times10^6$ & $0.05$ & TNG300-1 \\
3.4 & full-sky 4 & continuous & HEALPix & random & MWE 4 & $3\times10^6$ & $0.05$ & TNG300-1 \\
3.5 & full-sky 5 & continuous & HEALPix & random & MWE 5 & $3\times10^6$ & $0.05$ & TNG300-1 \\
3.6 & full-sky 6 & continuous & HEALPix & random & MWE 6 & $3\times10^6$ & $0.05$ & TNG300-1 \\
4 & pencil-beam & continuous & pencil beam & random & MWE 1 & $1\times10^6$ & $0.05 $ & TNG300-1 \\
5.1 & halo mass 0.1 & continuous & uniform & MH & random & $1\times10^5$ & $0.1$ & TNG300-1 \\
5.2 & halo mass 0.5 & continuous & uniform & MH & random & $1\times10^5$ & $0.5$ & TNG300-1 \\
5.3 & halo mass 1.0 & continuous & uniform & MH & random & $1\times10^5$ & $1.0$ & TNG300-1 \\
5.4 & halo mass 1.5 & continuous & uniform & MH & random & $1\times10^5$ & $1.5$ & TNG300-1 \\
5.5 & halo SFR 0.1 & continuous & uniform & SFH & random & $1\times10^5$ & $0.1$ & TNG300-1 \\
5.6 & halo SFR 0.5 & continuous & uniform & SFH & random & $1\times10^5$ & $0.5$ & TNG300-1 \\
5.7 & halo SFR 1.0 & continuous & uniform & SFH & random & $1\times10^5$ & $1.0$ & TNG300-1 \\
5.8 & halo SFR 1.5 & continuous & uniform & SFH & random & $1\times10^5$ & $1.5$ & TNG300-1 \\
6.1 & stacked & stacked & $\hat{e}_x, \,\hat{e}_y, \,\hat{e}_z$ & random & random & $1.5\times10^5$ & $5.5$ & TNG300-1 \\
6.2 & medium & stacked & $\hat{e}_x, \,\hat{e}_y, \,\hat{e}_z$ & random & random & $1.5\times10^5$ & $5.5$ & TNG300-2 \\
6.3 & low & stacked & $\hat{e}_x, \,\hat{e}_y, \,\hat{e}_z$ & random & random & $1.5\times10^5$ & $5.5$ & TNG300-3 \\
6.4 & midsize & stacked & $\hat{e}_x, \,\hat{e}_y, \,\hat{e}_z$ & random & random & $1.5\times10^5$ & $5.5$ & TNG100-2 \\
6.5 & small & stacked & $\hat{e}_x, \,\hat{e}_y, \,\hat{e}_z$ & random & random & $1.5\times10^5$ & $5.5$ & TNG50-3 \\
7 & trapezoidal & trapezoidal & $\hat{e}_x, \,\hat{e}_y, \,\hat{e}_z$ & random & random & $1.5\times10^5$ & $5.5$ & TNG300-1 \\
8 & periodic & periodic & $\hat{e}_x, \,\hat{e}_y, \,\hat{e}_z$ & random & random & $1.5\times10^5$ & $5.5$ & TNG300-1 \\
\enddata
\tablecomments{\textbf{Summary table of all presented DM catalogs.} For each catalog, we report the name, underlying method, direction of the rays, ray starting points, observer location, number of rays, maximum integration redshift, and underlying IllustrisTNG simulation.
The four methods we employ (continuous, stacked, periodic, and trapezoidal) are described in detail in Section \ref{subsec:raytracing} and Figure \ref{fig:comp_algorithms}.
When we say that we use rays with a ray direction vector of $\vec{n}=(1,5,25)$, we in fact use twelve different signed permutations of $\vec{n}$.
This is necessary to correctly model cosmic variance since choosing a single ray direction also implies selecting a single realization of cosmic modes.
We find that taking only a single ray direction can lead to a change in the distribution of up to $2$-$5\%$.
The different start points we considered, ranging from massive halos (MH), star-forming halos (SFH), and random locations, are outlined in Section \ref{subsec:startpoints_halo}.
The location of the observer, chosen to be either within a Milky Way-like environment (MWE), within a plane perpendicular to the ray vector, or at a random location, is described in Sections \ref{subsec:choicerayvector} and \ref{subsec:observer_position}.
In the case of the HEALPix projections, we use an $N_{\mathrm{side}}$ of $512$, corresponding to approximately $3\times10^6$ rays.
For the stacked and trapezoidal methods, the number of rays refers to the lines of sight sent through the simulation.
In general, according to the definition of these two methods, which is based on the convolution of distributions (see Section \ref{subsec:raytracing}), one can construct a number of observations much larger than the number of rays.
Each catalog listed in this table includes DMs from redshift $0$ to a maximum redshift, with a redshift step size of $0.01$, except for the trapezoidal catalog, which is based on a redshift step size of around $0.1$.
For details about the underlying simulation from IllustrisTNG see Sections \ref{subsec:IllustrisTNG} and \ref{subsec:size_resoluttion}.
As part of this work, we make all DM catalogs publicly available.
The catalogs can be downloaded at \href{https://ralfkonietzka.github.io/fast-radio-bursts/ray-tracing-catalogs/}{https://ralfkonietzka.github.io/fast-radio-bursts/ray-tracing-catalogs/}.
}
\end{deluxetable*}

\section{Results and Discussion}\label{sec:Results}
In this section we present our final results, summarized in over $20$ DM catalogs.
The catalogs differ in the underlying method (continuous, stacked, periodic, and trapezoidal), the ray direction and number, the location of the starting points, the location of the observer, the maximum redshift, as well as the simulation boxsize and resolution of the simulation.
We summarize and compare all catalogs to each other in Table \ref{tab:DMcatalogs}.
The labels in the first column correspond to a subdivision of the catalogs into 8 categories:  the continuous catalog (1), the random catalog (2), the full-sky catalogs (3.x), the pencil-beam catalog (4), the halo catalogs (5.x), the catalogs based on the stacked method (6.x), the trapezoidal catalog (7), and the periodic catalog (8).
In the following sections we will make use of these catalogs to highlight different results and applications.

\subsection{Effect and Choice of the Ray-tracing Method}\label{subsec:method_comparison}
In Section \ref{sec:SimulatingDM} we introduced four different methods for generating DM catalogs in cosmological simulations.
As shown in Figure \ref{fig:comp_algorithms}, the continuous method sends rays through the simulation volume under a given angle with respect to the box axes.
The exact direction of the rays is an important choice, which we discuss in Section \ref{subsec:choicerayvector}.
For the main continuous catalog in this paper, described in Section \ref{subsec:continuousmethod} we use $\vec{n} = (1, 5, 25)$ with twelve different signed permutations.
In contrast, the periodic method is equivalent to keeping the direction vector equal to one of the box axes (see Section \ref{subsec:periodicmethod}).
While applying the periodic method makes use of the simulations periodic boundary conditions, the rays also travel periodically repeatedly through the same path of the simulation.
Avoiding these repetitions and breaking the continuity between two simulation boxes at the same redshift by randomly rotating and shifting the boxes together with randomly sampling the rays leads to the stacked method, which we show in orange in Figure \ref{fig:comp_algorithms} and describe in Section \ref{subsec:stackedmethod}.
Leaving gaps between the simulated volumes and accounting for these gaps by artificially stretching the distributions results in the trapezoidal method (see Section \ref{subsec:trapeziodalmethod}, red color in Figure \ref{fig:comp_algorithms}).

For all four methods we output the DM distribution $p(\mathrm{DM}|z)$ as a function of redshift $z$.
For the continuous, stacked, and periodic methods, we generate catalogs with a step size in redshift of 0.01.
By construction, for the trapezoidal approach we can only output distributions at redshifts equal to the redshifts of the underlying simulation boxes, which implies a step size in redshift of 0.1 at small redshifts ($z \ll 1$) and of 1.0 at large redshifts ($z \gg 1$). This is the reason why there are fewer data points for the trapezoidal method in all figures compared to the other approaches.
In all comparisons, we treat the continuous method as the ground truth.

\begin{figure}[]
    \centering
    \includegraphics[width=0.47\textwidth]{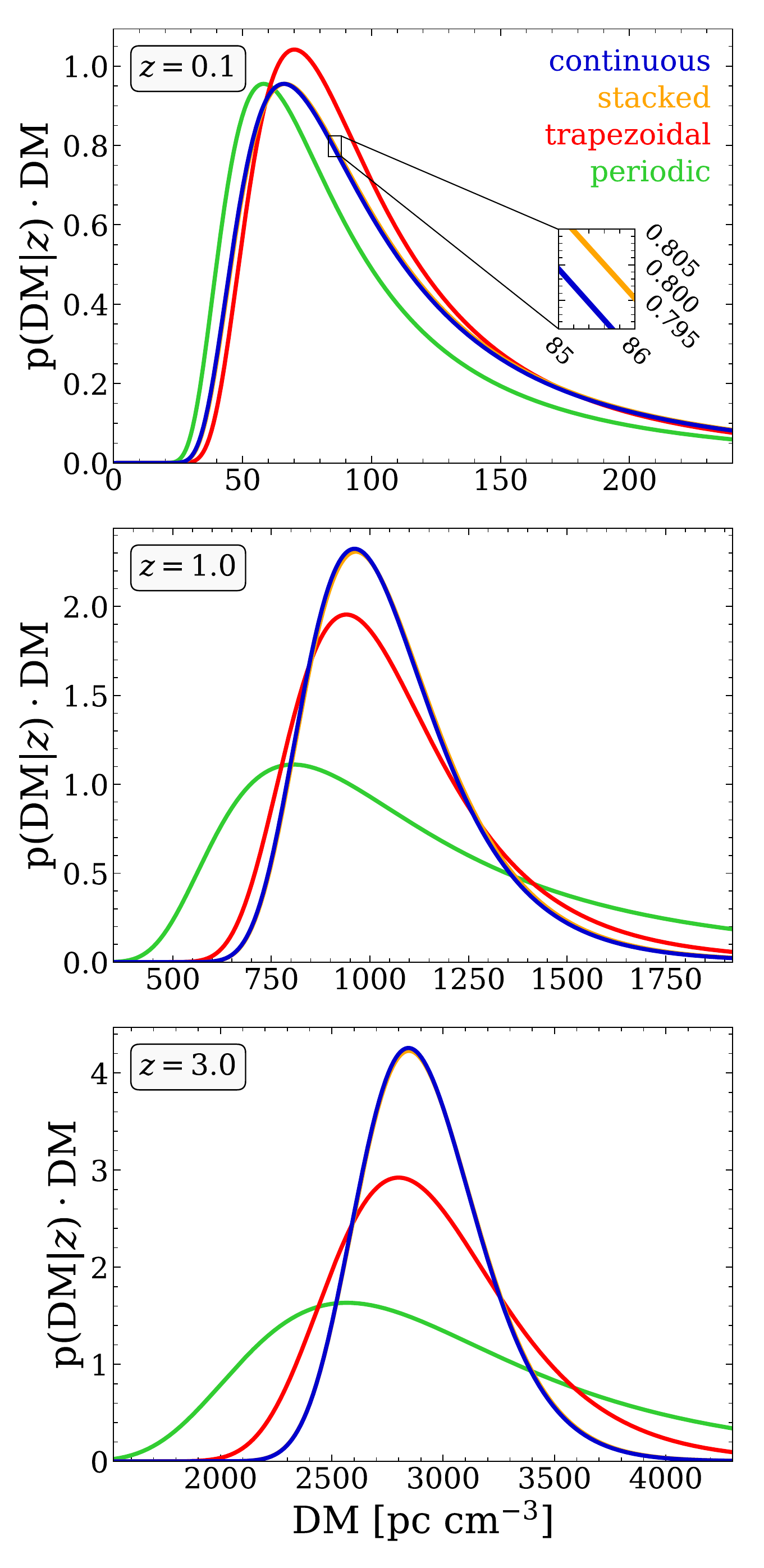}
    \caption{\textbf{Comparison between PDFs resulting from different ray-tracing methods}.
    For redshifts $0.1$, $1.0$ and $3.0$ we show the PDFs resulting from the four different ray-tracing methods discussed in this work.  The continuous approach is displayed in blue, the stacked ansatz in orange, and the periodic and trapezoidal methods in green and red, respectively.
    The colors are the same as in Figure \ref{fig:comp_algorithms}.
    We see that at all redshifts the continuous and stacked approaches produce consistent results, while the periodic and trapezoidal methods distort the distribution.
    We show a more detailed analysis of the differences between the methods using the first four moments of $p(\mathrm{DM}|z)$ in Figure \ref{fig:comp_mean_sigma_methods}.
    In all panels the data was fitted to the modified normal density function $p_{np,31}$ (see Section \ref{subsec:shape_DM}).
    \label{fig:comp_methods_hist}}
\end{figure}

We compare the distributions of all four methods to each other at redshifts $0.1$, $1.0$, and $3.0$ in Figure \ref{fig:comp_methods_hist}.
We see that at all three redshifts the continuous and stacked approaches are consistent.
However, already at redshift of 0.1, the trapezoidal and periodic approximations clearly distort the distribution.
The distortion becomes worse at higher redshifts.

To compare the differences in the distributions in more detail, we can analyze their moments, defined by
\begin{equation}\label{eq:4moments}
    \left<x^n\right>_z = \int_{DM}\,x^n\,p(x|z)\,dx,
\end{equation}
where $n \geq 1$ denotes the order of the moment, such that we obtain the mean $\mu$ for $n = 1$ and the variance $\sigma^2$ using $n = 2$ (modulo the square of the mean of $p(x|z)$).

In this work, we will also analyze the third ($n=3$) and fourth ($n=4$) moment, the skewness $\gamma$ and the kurtosis $\kappa$.
We define the skewness $\gamma$ as
\begin{equation}\label{eq:skewness}
    \gamma = \frac{\left<x^3\right>_z - 3\,\mu\,\sigma^2 + \mu^3 }{\sigma^{3}}
\end{equation}
and the kurtosis $\kappa$ as
\begin{equation}\label{eq:kurtosis}
    \kappa = \frac{\left<x^4\right>_z - 4\,\gamma\,\mu\,\sigma^3 - 6\,\mu^2\,\sigma^2 - \mu^4 - 3\,\sigma^4}{\sigma^{4}},
\end{equation}
such that both $\gamma$ and $\kappa$ are equal to $0$ for a normal distribution.
A distribution with $\gamma > 0$ is right-skewed, while a distribution with $\gamma < 0$ is left-skewed.
Similarly, a distribution with $\kappa > 0$ is called leptokurtic and has a higher peak and wider tails than a normal distribution. In contrast, a platykurtic distribution corresponds to $\kappa < 0$ and has a smaller peak and narrower tails than a normal distribution.

\begin{figure*}[]
    \centering
    \includegraphics[width=\textwidth]{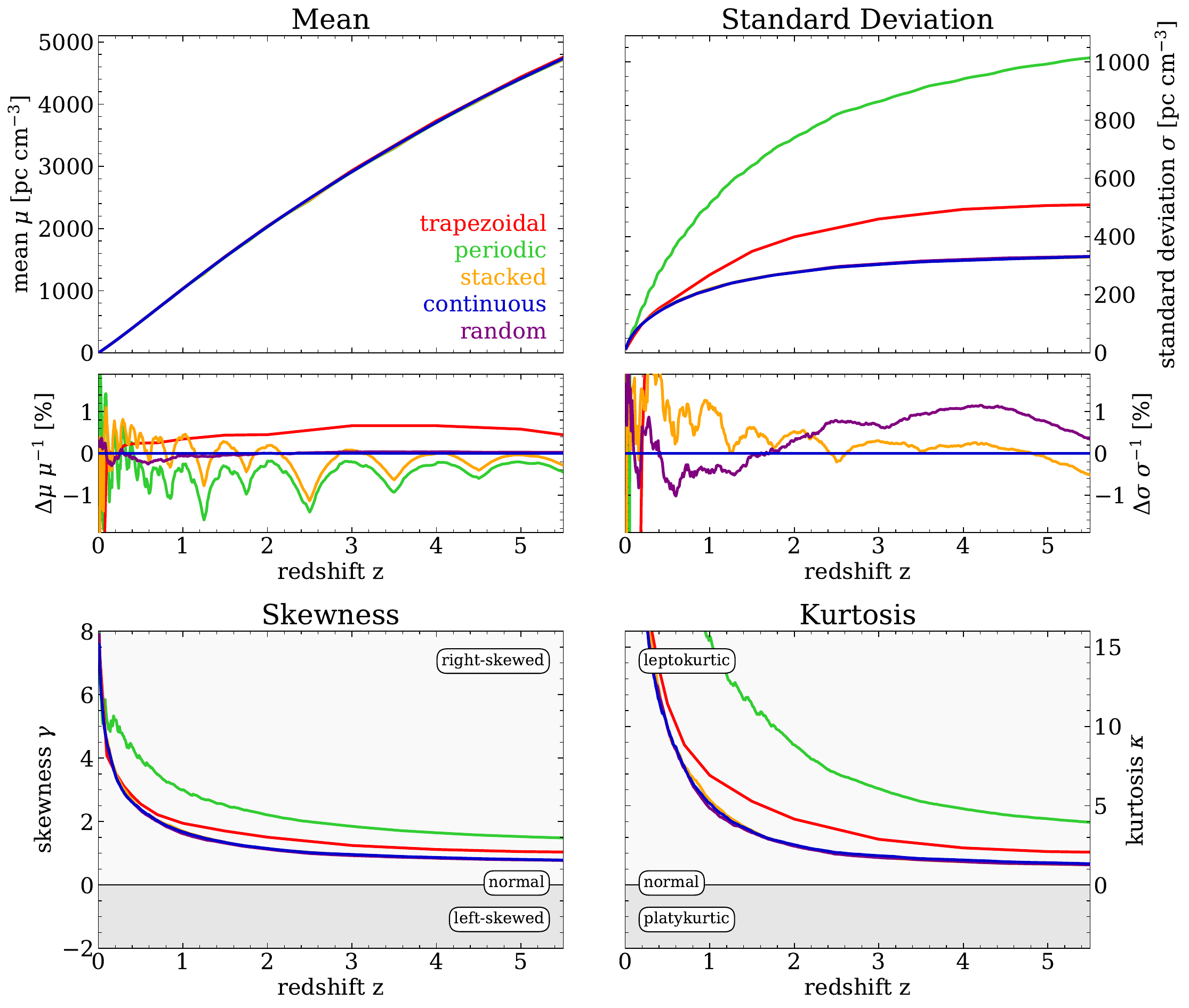}
    \caption{\textbf{Evolution of the first four moments of the DM distribution $\mathbf{p(DM|z)}$ for different ray-tracing methods}. We show the mean (left upper panel), the standard deviation (right upper panel), the skewness (left lower panel) and the kurtosis (right lower panel) to visualize the imprint of all four ray tracing methods (continuous, stacked, periodic and trapezoidal) on the final DM distribution.
    The colors correspond to those used in the Figures \ref{fig:comp_algorithms} and \ref{fig:comp_methods_hist}.
    We also show in purple the random catalog that evolves rays in all directions on the sky.
    We see that all four methods are consistent with respect to their mean evolution.
    As expected, for the stacked and periodic catalog we see in the residuals the imprint of the simulation boxes used in the form of periodically appearing patterns that peak when we switch from simulation boxes at a given redshift to the next following redshift.
    The methods lead to different results when we examine higher moments.
    We see that the trapezoidal approach changes the standard deviation by over $50\%$ compared to the continuous catalog. 
    The skewness and kurtosis are affected by over $30\%$ and $60\%$ respectively.
    The effect of the periodic approach is even stronger.
    The standard deviation is biased by over $200\%$, the skewness by over $90\%$ and the kurtosis by over $200 \%$.
    The stacked and random catalog are consistent with the continuous method across all moments.
    \label{fig:comp_mean_sigma_methods}}
\end{figure*}

Very massive clusters ($>10^{14.5}\,M_{\odot}$) are rare in TNG due to the finite simulation volume \citep{Nelson_2024_TNGcluster}.
We expect these clusters to dominate the high DM imprint into $p(\mathrm{DM}|z)$ at a given redshift $z$ (see the red clumps in Figures \ref{fig:fullskymap} and \ref{fig:fullsky_observer_change}, and for example, \cite{Walker_2024_DMcatalog}).
To avoid potential biases introduced by this constraint, we trim all moments by $0.1\%$ at the high DM end for all reported numbers.
We vary this choice between $0\%$ and $1\%$, finding no effect on our results.
We also explore using quantiles instead of moments and again find that the results are the same.

We present all four moments using all four ray-tracing methods (continuous, stacked, periodic and trapezoidal) in Figure \ref{fig:comp_mean_sigma_methods}.
As expected, we see that the overall mean evolution of the distribution $p(\mathrm{DM}|z)$ is well reproduced by all methods.

The continuous and stacked approaches give nearly identical results both visually (Figure~\ref{fig:comp_methods_hist}) and quantitatively when comparing their mean, standard deviation, skewness, and kurtosis (Figure~\ref{fig:comp_mean_sigma_methods}).
Only in the residuals do we see that the discontinuities in the stacked catalog introduced at the boundary between two simulation volumes lead to a small difference between the two methods.
As these differences are on the percent level, we conclude that the continuous and stacked methods are consistent.
We emphasize that we prefer the continuous approach since it avoids discontinuities in the transition from one simulation volume to another.

In contrast, as expected from the histograms in Figure \ref{fig:comp_methods_hist}, the trapezoidal and periodic methods show significant differences in all higher moments (standard deviation, skewness, and kurtosis) compared to the continuous method.
Figure~\ref{fig:comp_mean_sigma_methods} shows that the standard deviation between the trapezoidal and continuous methods differs by over $50\%$.
Comparing the periodic to the continuous approach results in a difference in the standard deviation of over $200\%$.
We also find that the trapezoidal and periodic methods skew the distribution more to the right compared to our ground truth by over $30\%$ and $90\%$, respectively.
In addition, $p(\mathrm{DM}|z)$ appears to be more leptokurtic by over around $60\%$ and $200\%$ compared to the continuous ansatz when relying on the trapezoidal and periodic methods.

We conclude that the continuous and stacked approaches are two valid approaches to calculate $p(\mathrm{DM}|z)$ in numerical simulations.
We will use the continuous and stacked methods in all further analyses.
In contrast, the trapezoidal and periodic approaches significantly distort the distributions.
This distortion occurs because neither method correctly accounts for cosmic variance.
As we explained in Section $\ref{sec:SimulatingDM}$, in both cases, the distributions are stretched to reproduce the correct mean, which changes the width, inclination, and peakiness of the distribution.

In addition, to the four different ray-tracing methods we also show in Figure \ref{fig:comp_mean_sigma_methods} the random catalog in purple.
This approach starts at random positions in the simulation and evolves under random direction vectors $\vec{n}_i$, covering the entire sky.
We find that the random catalog does not differ from the continuous catalog which includes rays that all evolve along the same direction $\vec{n}=(1, 5, 25)$ (module $12$ signed permutations).
This bolsters our discussion in Section \ref{subsec:choicerayvector} where we argued that the full-sky catalog is consistent with the plane-parallel approach.

\begin{figure}[h]
    \centering
    \includegraphics[width=0.47\textwidth]{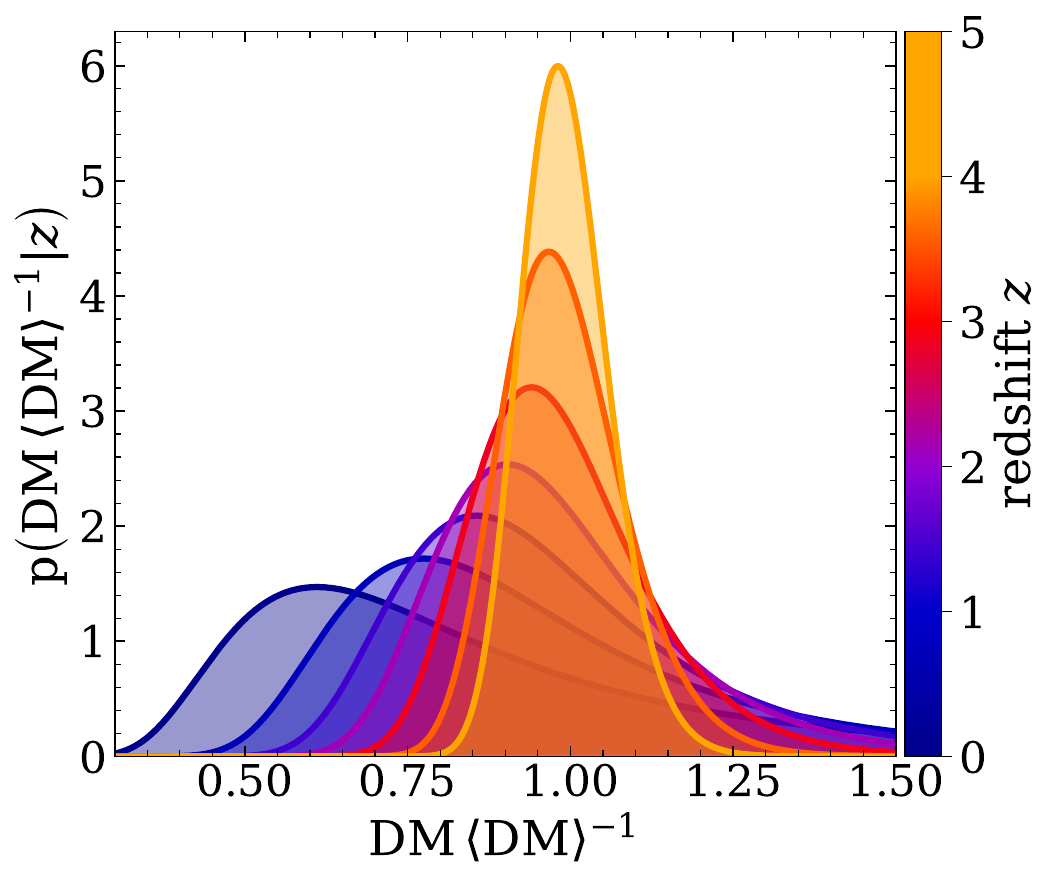}
    \caption{\textbf{Evolution of $\mathbf{p(DM \, \left<DM\right>^{-1}|z)}$.} We show the evolution of the probability density function $p(\mathrm{DM} \, \left<\mathrm{DM}\right>^{-1}|z)$ at several redshifts after fitting to the data the modified-normal $p_{np,31}$ distribution (see Section \ref{subsec:shape_DM}).
    The results from the continuous catalog were used (see Section \ref{subsec:method_comparison}).
    \label{fig:hist_evolution}}
\end{figure}

Figure \ref{fig:comp_mean_sigma_methods} allows us not only to compare the different methods, but also to examine the shape of $p(\mathrm{DM}|z)$.
We observe that the skewness and kurtosis approach zero at higher redshifts, implying that $p(\mathrm{DM}|z)$ becomes increasingly normally distributed.
To emphasize this evolution of the probability density function $p(\mathrm{DM}|z)$, we show in Figure \ref{fig:hist_evolution} the PDF after normalizing the DM with its mean $\left<\mathrm{DM}\right>$ at each redshift.

\subsection{The Shape of the Dispersion Measure Distribution}\label{subsec:shape_DM}
In this section we explore which functional form best describes the shape of the DM distribution $p(\mathrm{DM}|z)$ at a given redshift $z$.
As we pointed out earlier, the distribution is right-skewed and leptokurtic across all redshifts between 0 and 5.5.
This is the motivation for the following six functional forms that we will analyse in more detail.

\begin{figure*}[]
    \raggedright
    \includegraphics[width=\textwidth]{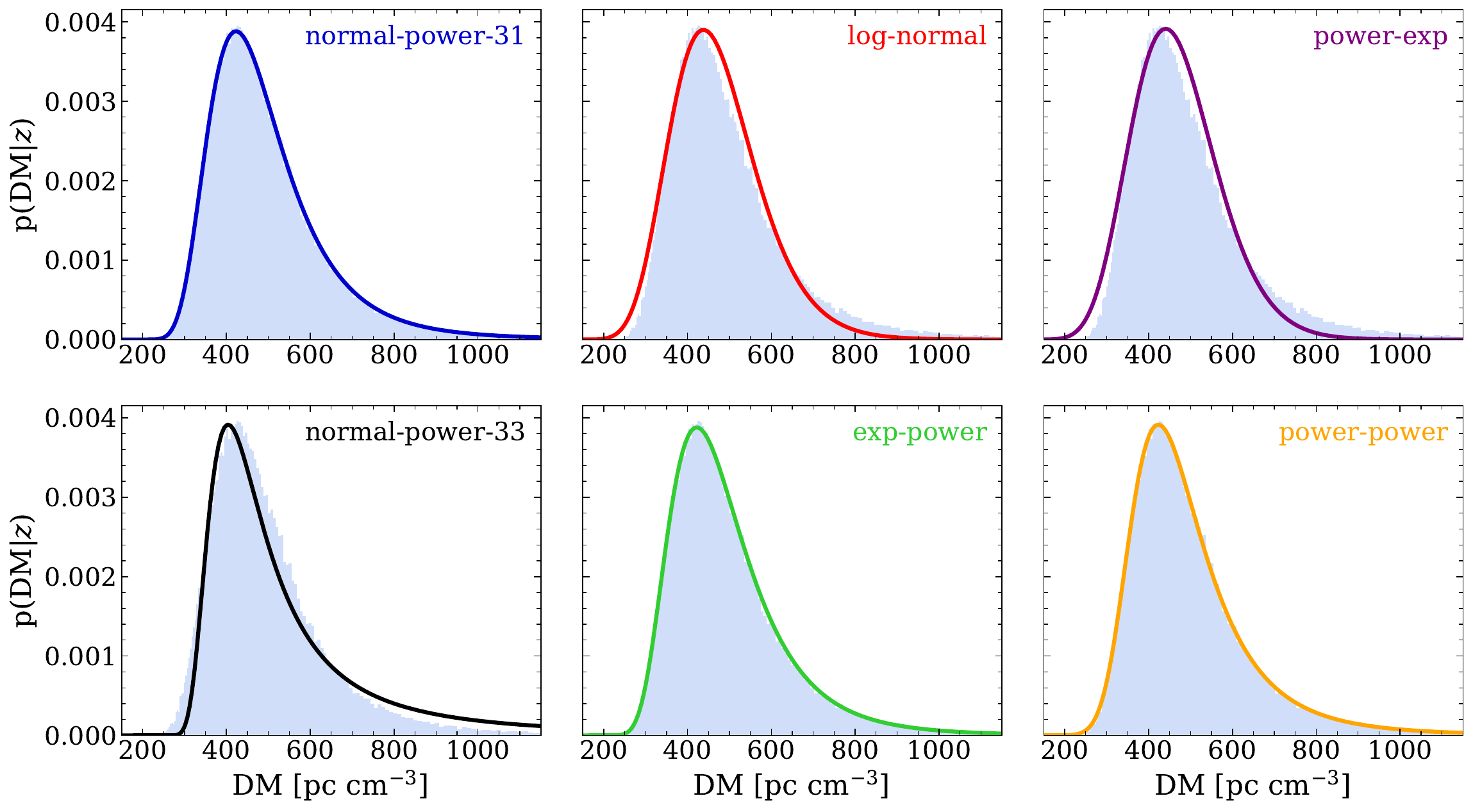}
    \caption{\textbf{Best model for $\mathbf{p(DM|z)}$.}
    We show $p(\mathrm{DM}|z)$ in light blue, computed using the continuous catalog, and compare this distribution to all six proposed analytic distributions ($p_{ep}$, $p_{pe}$, $p_{pp}$, $p_{ln}$, $p_{np,33}$, $p_{np,31}$) at redshift $z=0.5$.
    We see that the high DM tail of the distribution is best described by a power-law drop off.
    On the low DM left side of the distribution an exponential rise describes the data best.
    The modified-normal distribution with a power-law drop-off ($p_{np,31}$) fits the data very well.
    The power-exponential distribution $p_{pe}$ is not well matched to the DM data.
    The power-power $p_{pp}$ distribution and power-exponential distribution $p_{pe}$ work well at this redshift, but they break down at higher redshifts (see Figure \ref{fig:KS_stats_best_fit}).
    The normal-power distribution $p_{np,33}$ and the log-normal distribution $p_{ln}$ are not well matched to the data.
    Both distributions misestimate the low DM part of the distribution and are unable to fit the long DM tail. \label{fig:best_model}}
\end{figure*}

\begin{figure}[]
    \centering
    \includegraphics[width=0.47\textwidth]{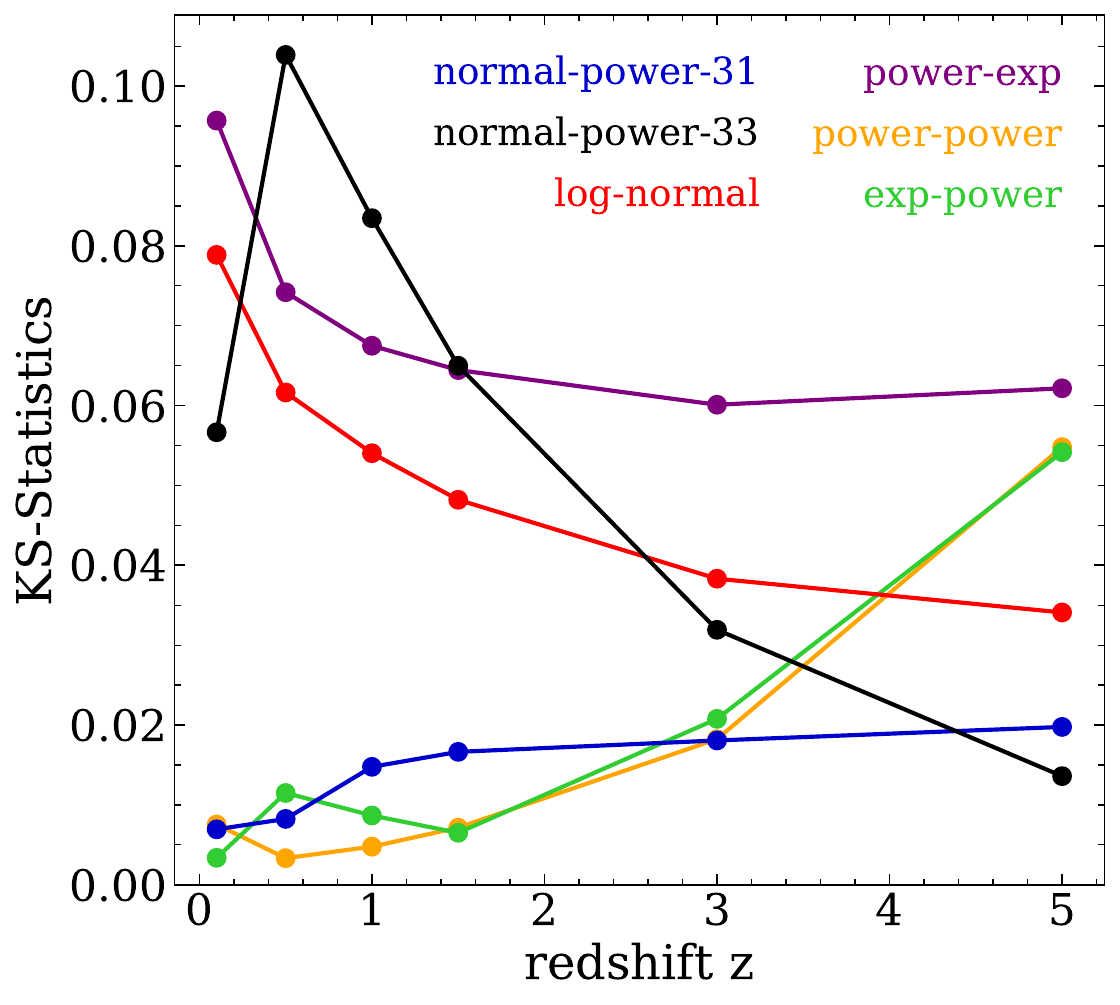}
    \caption{\textbf{Kolmogorov–Smirnov (KS) statistics.}
    We compare all six proposed distributions ($p_{ep}$, $p_{pe}$, $p_{pp}$, $p_{ln}$, $p_{np,33}$, $p_{np,31}$) across six redshift slices at redshifts $0.1$, $0.5$, $1.0$, $1.5$, $3.0$ and $5.0$ using the Kolmogorov–Smirnov (KS) statistics.
    Of all the models, the modified-normal distribution with a power-law drop-off ($p_{np,31}$) is the only one that fits the data well across all redshifts.
    \label{fig:KS_stats_best_fit}}
\end{figure}

First, we introduce a probability density function $p_{ep}$ that includes an exponential rise for small DM and a power law drop for large DM, defined with the parameters $\alpha>0$, $\beta>1$ and $\mu>0$:
\begin{align}\label{eq:pdf_ep}
    p_{ep}(x, \alpha, \beta, \mu) &= c\, \exp\left(-\left(\frac{\mu}{x}\right)^\alpha\right)\,\left(\frac{\mu}{x}\right)^\beta\\
    c^{-1}  &= \frac{\mu}{\alpha} \, \Gamma\left(\frac{\beta-1}{\alpha}\right),
\end{align}
where $\Gamma(\cdot)$ denotes the gamma function.

Second, we will use a probability density function $p_{pe}$, that rises according to a power law at small DM and decreases exponentially for large DM, with the parameters $\alpha > 0$, $\beta > 0$ and $\mu >0$:
\begin{align}\label{eq:pdf_pe}
    p_{pe}(x, \alpha, \beta, \mu) &= c\,\left(\frac{x}{\mu}\right)^\alpha\exp\left(-\left(\frac{x}{\mu}\right)^\beta\right)\\
    c^{-1}  &= \frac{\mu}{\beta} \, \Gamma\left(\frac{\alpha+1}{\beta}\right).
\end{align}

Third, we consider a PDF that increases and decreases according to a power law with the free parameters $\alpha>0$, $\beta>0$, $k > \frac{\alpha+1}{\beta}$ and $\mu>0$:
\begin{align}\label{eq:pdf_pp}
    p_{pp}(x, \alpha, \beta, k, \mu) &= c\,\left(\frac{x}{\mu}\right)^\alpha\,\left(1 + \left(\frac{x}{\mu}\right)^\beta\right)^{-k}\\
    c^{-1}  &= \frac{\mu}{\beta} \, \mathrm{B}\left(\frac{\alpha+1}{\beta},\,k-\frac{\alpha+1}{\beta}\right),
\end{align}
where $B(\cdot,\,\cdot)$ denotes the beta function.
With these three PDFs we cover a variety of different right-skewed functional forms, including the Weibull, generalized gamma, chi-squared, log-logistic, Nakagami, Erlang, inverse Beta, Dagum and Burr distributions.  

In addition to that, we investigate the log-normal distribution $p_{ln}$, which can be defined with the mean $\mu$ and the variance $\sigma^2$ as
\begin{equation}\label{eq:pdf_ln}
    p_{ln}(x, \mu, \sigma) = \frac{1}{x\,\sigma\,\sqrt{2\pi}}\,\exp\left(-\frac{\left(\ln(x) - \mu\right)^2}{2\,\sigma^2}\right).
\end{equation}

The fifth and sixth distributions that we consider are described by the following modified-normal density function with a power law tail at high DM and free parameters $\mu$ and $\sigma$:
\begin{align}\label{eq:pdf_np}
    p_{np}(x, \mu, \sigma) = c\,\exp\left(-\frac{\mu^2\left(x^{-\alpha}\,\mu^\alpha - 1\right)^2}{2\,\sigma^2\,\alpha^2}\right)\left(\frac{\mu}{x}\right)^{\beta}\\
    c^{-1} = \frac{\mu}{\alpha}\,e^{-\frac{\eta^2}{2}}\,\left(\frac{\alpha\,\sigma}{\mu}\right)^\delta\,\Gamma\left(\delta\right)\,\mathrm{D}_{-\delta}\left(-\sqrt{2}\,\eta\right),
\end{align}
where $\delta = \frac{\beta-1}{\alpha}$, $\eta^2 = \frac{\mu^2}{2\,\sigma^2\,\alpha^2}$ and $D_\cdot(\cdot)$ is the parabolic cylinder function.
Note that in equation (\ref{eq:pdf_np}), in contrast to the previous PDFs, the parameters $\alpha$ and $\beta$ are fixed and not changing with redshift.
We allow for two different choices of $\alpha>0$ and $\beta>1$ that are motivated below.

The distribution function $p_{np}$ is very similar to the functional form suggested by \cite{Miralda_2000_NormalFit} and \cite{Macquart_2020} with the difference of the definition of $\sigma$ and that in the literature $\mu$ is forced to be equal to the mean of the distribution.
\cite{Macquart_2020} sets $\alpha$ and $\beta$ equal to $3$.
We also use this choice with our functional form and refer to this as $p_{np,33}$.

The second choice of $\alpha$ and $\beta$ is obtained by optimizing for these parameters across six redshift slices at redshifts $0.1$, $0.5$, $1.0$, $1.5$, $3.0$ and $5.0$ using the nested sampling code \textsc{dynesty} \citep{Speagle_2020_dynesty}.
We adopt pairwise independent uniform priors on each parameter, and run a combination of random-walk sampling with multi-ellipsoid decompositions and $1000$ live points. 
We find that $\alpha \approx 1$ and $\beta \approx 3.3$ describe the data best, while we caution that the binning of the data and the choice and weighting of the slices can modify the parameters by about $0.5$.
A more detailed analysis of the parameters $\alpha$ and $\beta$ that involves a physical interpretation of the parameters is beyond the scope of this work.
However, we note that the exact values of $\alpha$ and $\beta$ are likely to encode the amount of electrons in halos as well as the shape of the densest structures in the cosmic web.
We refer to the distribution $p_{np}$ with $\alpha=1$ and $\beta=3.3$ as $p_{np,31}$.

Furthermore, we note that for $\frac{\mu}{\sigma} \gg 1$ the distributions $p_{np,33}$ and $p_{np,31}$ become normally distributed with mean $\mu$ and variance $\sigma^2$.
However, in general $\mu$ is not equal to the mean of the distributions.
As we show in Figure \ref{fig:hist_evolution} and pointed out in Section \ref{subsec:method_comparison}, for large redshifts $z$, also the DM distribution function becomes increasingly normally distributed, implying that the last three described functional forms ($p_{ln}$, $p_{np,33}$ and $p_{np,31}$) are particularly interesting for us as they incorporate this behavior.

To find the best-fit parameters of all six proposed distributions ($p_{ep}$, $p_{pe}$, $p_{pp}$, $p_{ln}$, $p_{np,33}$, $p_{np,31}$) across six redshift slices at redshifts $0.1$, $0.5$, $1.0$, $1.5$, $3.0$ and $5.0$ we again use the nested sampling code \textsc{dynesty}. We run a combination of random-walk sampling with multi-ellipsoid decompositions and $1000$ live points based on pairwise independent uniform priors on each parameter.
We show the best-fits at redshift $0.5$ in Figure \ref{fig:best_model}.

In Figure \ref{fig:KS_stats_best_fit} we compare the distributions across all six redshift slices using the Kolmogorov–Smirnov (KS) statistics defined via
\begin{equation}\label{eq:KS_stats}
    \mathrm{KS}(p) = \sup_x\left|F_p(x)-F_{DM}\right|,
\end{equation}
where $F_p(x)$ and $F_{DM}$ are cumulative distribution functions (CDFs) corresponding to the probability density function $p$ and the simulated DM distribution, respectively.

We find that the high DM tail of the distribution is best described by a power-law decline.
On the low DM, left side of the distribution an exponential rise describes the data best.
Consequently, $p_{pe}$ is rejected by the DM data.
The $p_{pp}$ distribution, describing the high DM tail correctly, provides a good fit at redshifts smaller than $3$ but fails at the high redshift end.
The $p_{ep}$ distribution incorporates the right functional form on both sides of the distribution and works well for redshifts smaller than $3$.
As expected, $p_{ep}$ and $p_{pp}$ break down at high $z$ as these forms do not approach a normal distribution for redshifts $z \gg 1$.

\begin{figure*}[]
    \centering
    \includegraphics[width=1.0\textwidth]{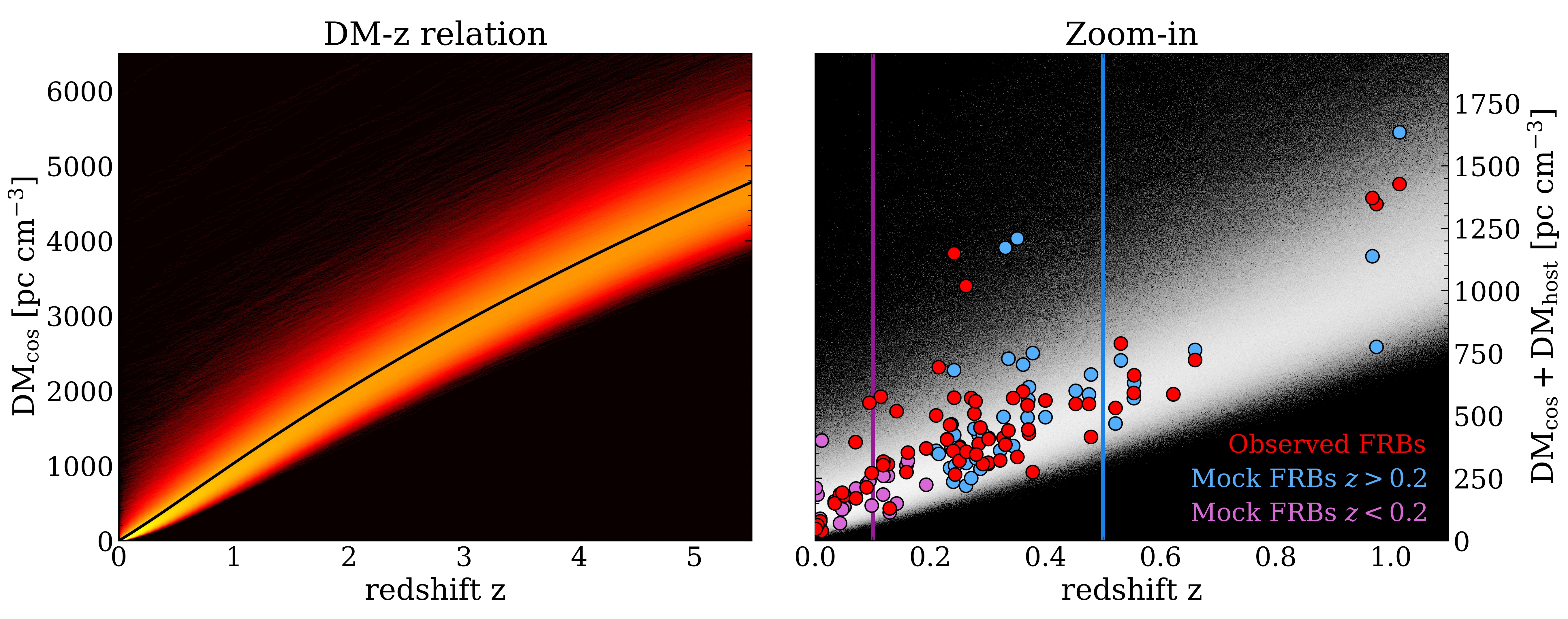}
    \caption{\textbf{The dispersion measure-redshift relation.}
    \textbf{Left panel.} We show the DM-z relation as a function of redshift $z$ using the continuous catalog from redshift $0$ to $5.5$.
    We see that for each redshift $z$ there is a minimum DM value that is at least reached.
    This sharp drop in $p(\mathrm{DM}|z)$ across all redshifts is often referred to as the DM cliff \citep{James_2022}.
    At the high DM end we observe that already at redshift $1$ some lines of sight can reach DM values of more than $3000\,$pc$\,$cm$^{-3}$, while the mean is around $1000\,$pc$\,$cm$^{-3}$.
    We show in black the analytic derivation of the mean DM evolution (see Section \ref{sec:Theoback}).
    \mbox{\textbf{Right panel.}} We zoom in on redshifts from $0$ to $1.25$ and overlay the distribution obtained from IllustrisTNG with an observed FRB DM sample from \cite{Connor_2024} in red.
    We see that the simulated data is in agreement with the observed FRBs.
    We also show mock FRBs drawn from one of our full-sky catalogs in purple for $z<0.2$ and blue $z>0.2$ (for details see Section \ref{subsec:DSAsampledata}).
    The vertical lines correspond to the redshifts of the full-sky maps displayed in the background of Figure \ref{fig:mollview_real_mock_data}.
    \label{fig:DM_z_relation}}
\end{figure*}

This high redshift behavior is included in $p_{np,33}$. 
However, the slopes of $\alpha = 3$ and $\beta =3$ introduced by \mbox{\cite{Macquart_2020}} are not well matched to the data across all redshifts.
To improve on this, \cite{Walker_2024_DMcatalog} and \cite{Sharma_2025} suggest using the log-normal distribution $p_{ln}$ instead.
However, as we see in Figures \ref{fig:best_model} and \ref{fig:KS_stats_best_fit}, the log-normal distribution $p_{ln}$ does not correctly fit the distribution.
The log-normal distribution overestimates the low DM part and is unable to fit the long DM tail.
The latter is particularly interesting for studies of the halo component of the \mbox{cosmic web}.

In contrast, the $p_{np,31}$ distribution is the only model that fits the data well at all redshifts.
Notably, among the analyzed models that tend to become normal at high redshifts, it is by far the best model.
Furthermore, the $p_{np,31}$ distribution stands out since it has only two free parameters, whereas the other distributions include up to four free parameters.

We conclude that the $p_{np,33}$ distribution and the log-normal distribution $p_{ln}$ should be avoided in future studies.
Instead, the use of $p_{np,31}$ with $\alpha\approx1$ and $\beta\approx3.3$ is recommended.
This distribution is also employed in all subsequent figures in this work.

\begin{figure}[]
    \raggedright
    \includegraphics[width=0.47\textwidth]{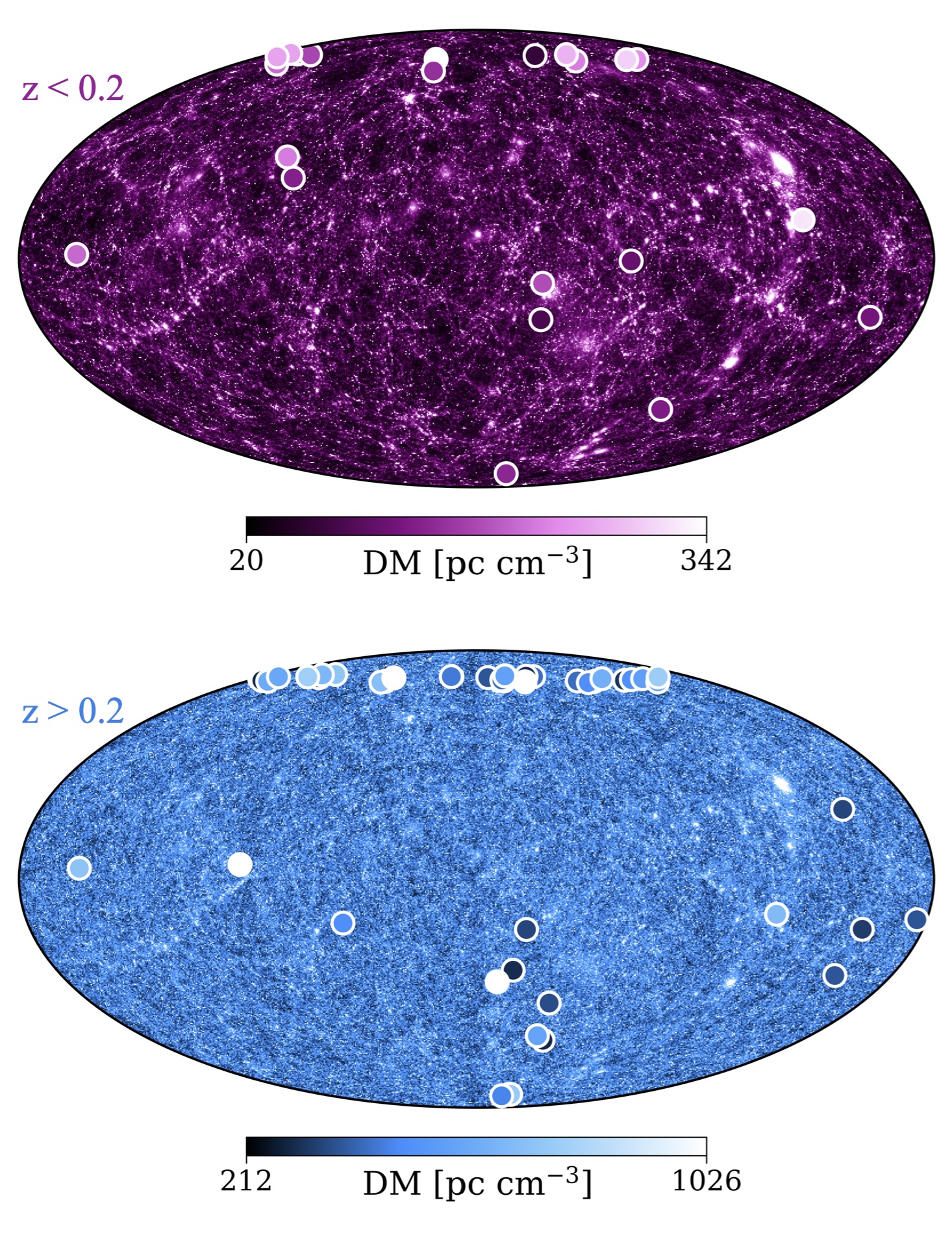}
    \caption{\textbf{FRB DMs on the full sky.}
    \textbf{Upper panel.} 
    We show our mock DM catalog derived from the full-sky catalog on the full sky, described in detail in Section \ref{subsec:DSAsampledata} at redshifts smaller than $0.2$.
    The points are colored according to DM.
    Since the observed DMs include the $\mathrm{DM}_{\mathrm{host}}$ contribution, we also add $\mathrm{DM}_{\mathrm{host}}$ to the simulated FRBs.
    The background map corresponds to the full-sky catalog at redshift $0.1$.
    \textbf{Lower panel.}
    We show our mock DM catalog derived from the full-sky catalog on the full sky, described in detail in Section \ref{subsec:DSAsampledata} at redshifts larger than $0.2$.
    The points are colored according to DM.
    As in the upper panel, we add $\mathrm{DM}_{\mathrm{host}}$ to the simulated FRBs.
    The background map corresponds to the full-sky catalog at redshift $0.5$.
    The vertical lines in Figure \ref{fig:DM_z_relation} correspond to the redshifts of the background maps used in the two panels.
    \label{fig:mollview_real_mock_data}}
\end{figure}

\subsection{The Dispersion Measure-Redshift Relation}\label{subsec:DM_z_relation}
With an accurate integration method in hand and a suitable functional form for $p(\mathrm{DM}|z)$, we now focus on the evolution of $p(\mathrm{DM}|z)$ with redshift.
To this end, we can examine the so-called dispersion measure-redshift (DM-z) relation, which we show in the left panel of Figure \ref{fig:DM_z_relation} using the continuous catalog from redshift \mbox{$0$ to $5.5$}.

We see that for each redshift $z$ there is a minimum DM value that is at least reached.
This sharp drop in $p(\mathrm{DM}|z)$ across all redshifts is often referred to as the DM cliff \citep{James_2022}.
At the high DM end we find that already at a redshift of $1$ some lines of sight can reach DM values of more than $3000\,$pc$\,$cm$^{-3}$, while the mean value is about $1000\,$pc$\,$cm$^{-3}$.
This shows that the crossing of very dense structures, mainly present in halos (see also Figures \ref{fig:fullskymap} and \ref{fig:fullsky_observer_change}), can lead to an extreme DM excess above the mean.

In the right panel of Figure \ref{fig:DM_z_relation}, we zoom-in into redshifts from $0$ to $1.25$ and overlay the relation obtained from IllustrisTNG with observed FRB DMs.
Our sample of FRBs \citep[see][and reference therein for details]{Connor_2024} includes observations from CHIME \citep{CHIME_2021, Marcote_2020_CHIME, Bhardwaj_2021_CHIME}, ASKAP \citep{McConnell_2016_ASKAP, Bannister_2017_ASKAP, Prochaska_2019_ASKAP}, Parkes \citep{Keith_2010_Parkes}, the DSA-110 \citep{Law_2024_DSA110}, Arecibo \citep{Arecibo_FRB_Spitler_2014}, FAST \citep{Niu_2022_FAST} and MeerKAT \citep{Caleb_2023_MeerKAT}.
We note that over half of the FRBs in this sample were detected by the DSA-110, followed by ASKAP and CHIME.

We subtract from the observational data the contribution of the Milky Way $\mathrm{DM}_{\mathrm{MW}}$ using a standard model from the literature \citep{Cordes_2002, Cordes_2003}.
As the observed FRB DMs include a host DM contribution, we also add to our modeled cosmic DM a host DM.
Based on recent observations \citep{Connor_2024} we model the host DM as a log-normal distribution with $\mu_{host} = 4.9$ and $\sigma_{host} = 0.56$. This is consistent with the host distribution found in the highest resolved IllustrisTNG run, TNG50-1 \citep{Theis_2024_host}.
We then draw random samples from the $p(\mathrm{DM}_{\mathrm{host}}|z)$ and our $p(\mathrm{DM}_{\mathrm{cos}}|z)$ that we plot as the white background in the right panel of Figure \ref{fig:DM_z_relation}.
Following from Section \ref{sec:Theoback}, $\mathrm{DM}_\mathrm{host}$ has to be weighted by a factor of $(1 + z)^{-1}$.
We find that the simulated and observed data are in good agreement with each other in terms of the mean and spread of the distributions.

In Figure \ref{fig:DM_z_relation} we also show mock FRBs drawn from one of our full-sky catalogs in purple for $z<0.2$ and blue $z>0.2$ (for details see Section \ref{subsec:DSAsampledata}).
These are the same mock FRBs as those displayed in Figure \ref{fig:mollview_real_mock_data}.

\begin{figure}[]
    \centering
    \includegraphics[width=0.43\textwidth]{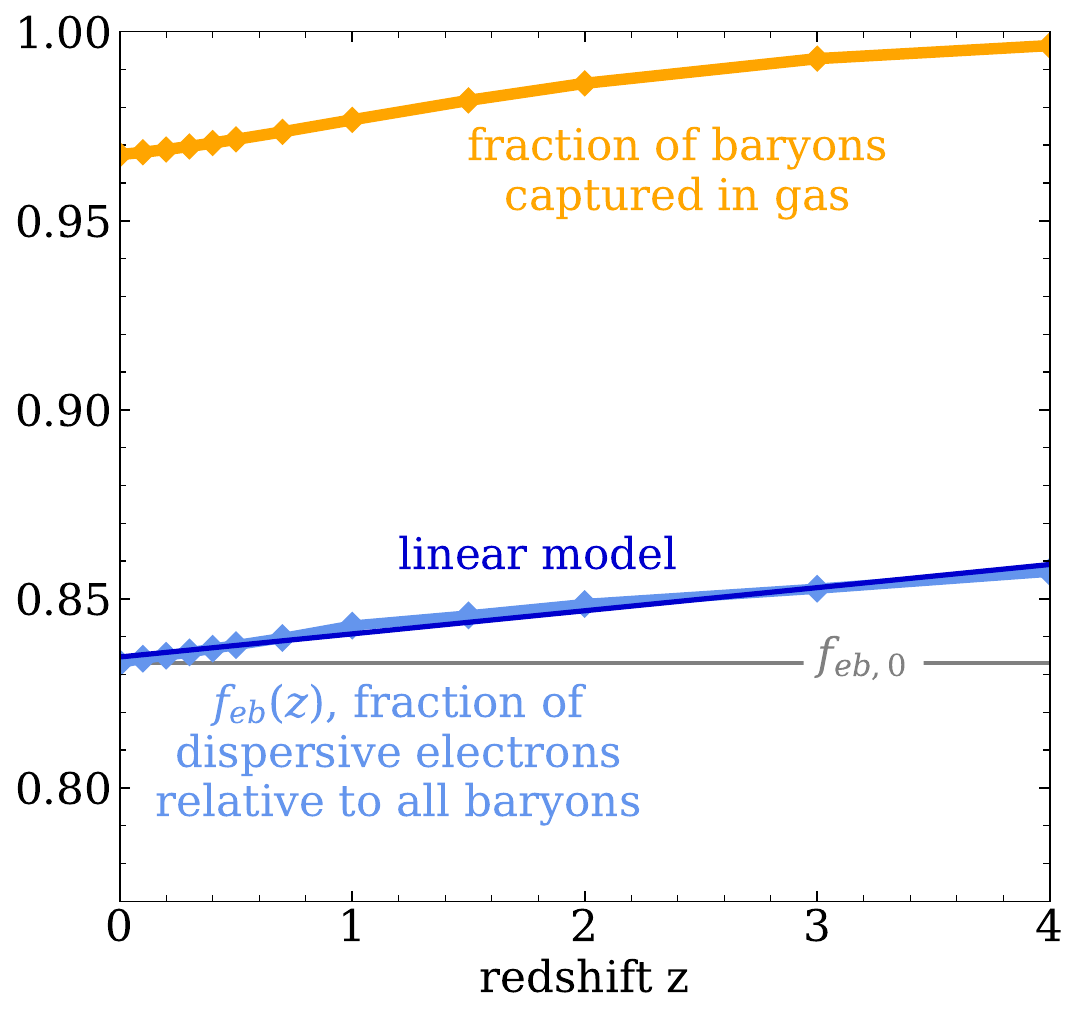}
    \caption{\textbf{Evolution of the fraction between the dispersive electron number density relative to the total baryon density in IllustrisTNG.}
    We show the fraction of the dispersive electron density relative to the total baryon density in IllustrisTNG, denoted by $f_{eb}(z)$, as a function of redshift in light blue.
    We see that $f_{eb}(z)$ increases roughly linearly with redshift, motivating us to fit the evolution of $f_{eb}(z)$ with a linear model.
    The linear fit is shown in dark blue.
    The zeroth order fit parameter $f_{eb,0}$, defined by $f_{eb}(z) \approx f_{eb,0} + f_{eb,1}\,z$ is highlighted with a gray horizontal line.
    In addition, we display the fraction of the gas mass relative to the total baryonic mass in IllustrisTNG.
    We see that at redshift $0$ the gas makes up more than $96\%$ of the baryonic matter.
    \label{fig:fraction_eb}}
\end{figure}

The evolution of the mean value of $p(\mathrm{DM}|z)$ can also be compared to the analytic formulation presented in Section \ref{sec:Theoback}.
To this end, we must first determine the behavior of the fraction of the dispersive electron density relative to the total baryon density $f_{eb}(z)$ in IllustrisTNG.
We show $f_{eb}(z)$ as a function of redshift $z$ in in Figure \ref{fig:fraction_eb}.
We see that $f_{eb}(z)$ is roughly linearly increasing, where we model $f_{eb}(z) \approx f_{eb,0} + f_{eb,1}\,z$.
At redshift $0$ the fraction $f_{eb}(0)=f_{eb,0}\approx0.83$.
Furthermore, we find that $f_{eb,1} \approx 0.006$.
In Figure \ref{fig:fraction_eb} the baryonic matter in the gas phase is also compared with the total baryonic mass to highlight that the fraction of baryons in stellar mass in IllustrisTNG is smaller than $4\%$.

Using the linear fit to the evolution of the fraction of the dispersive electron density compared to the total baryon density, $f_{eb}(z)$, we can calculate the integral defined in equation (\ref{eq:defDM3}) and compare it with the evolution of $p(\mathrm{DM}|z)$ derived from the continuous catalog.
We show the comparison in the left panel of Figure \ref{fig:DM_z_relation} as a black line.
We see that the analytical description perfectly describes the mean of the ray-tracing approach.

We note that the shape of the DM distribution is sensitive to the modeling of feedback and changes in the evolution of the UV background in IllustrisTNG.
In the context of the mean DM evolution, the effect of both, the feedback modeling and the UV background, are encapsulated in the function $f_{eb}(z)$.

Accordingly, we discuss how changes in $f_{eb}(z) \approx f_{eb,0} + f_{eb,1}\,z$ affect the mean of $p(\mathrm{DM}|z)$.
To this end, we separately modulate the zeroth order fraction $f_{eb,0}$ of $f_{eb}(z)$ and the first order fraction $f_{eb,1}$.
We find that a $10\%$ change in $f_{eb,0}$ affects the main slope of the DM-z relation by an amount equivalent to the overall variance of $p(\mathrm{DM}|z)$.
However, $f_{eb,0}$ is fully degenerate with the Hubble constant $H_0$ and the fraction of baryons in the Universe $\Omega_b$ (see equation \ref{eq:defDM3}).
Subsequently, any constraints on $f_{eb,0}$ (and with that also about the underlying feedback and the UV background) are limited by our knowledge of the underlying cosmology.
In contrast, the first-order parameter $f_{eb,1}$ is not directly degenerate with $H_0$ or $\Omega_b$, but $f_{eb,1}$ changes the overall slope of $p_z(DM)$ significantly less than $f_{eb,0}$.
We find that even a $200\%$ variation of $f_{eb,1}$ has a smaller effect than the $10\%$ modulation of $f_{eb,0}$.
We conclude that the mean of the DM-z relation can only be used to probe the ionization history of the Universe with prior knowledge of the cosmology.

Higher-order moments of $p(\mathrm{DM}|z)$ provide more insight into the effects of feedback and the evolution of the UV background.
The optimal ansatz for extracting this information is rerunning the same simulation setup with different initial conditions.
Such an analysis is beyond the scope of this work.
However, we note that the CAMELS project is a promising implementation of this approach \citep{CAMELS_presentation}.
Although constrained by a comparatively small box size (see Section \ref{subsec:size_resoluttion} for details), the CAMELS simulations offer a novel tool for studying FRBs under different initial conditions \citep[see for example,][]{Medlock_2024, Sharma_2025}.

\subsection{Simulated FRBs on the Full Sky}\label{subsec:DSAsampledata}
In this section we will now focus on the full-sky catalogs, one of which we show in Figure \ref{fig:fullskymap} together with the pencil beam catalog integrated up to redshift $z = 0.01$.
We see how the three-dimensional structure of the electron density is imprinted in the two-dimensional DM field.
Black cosmic voids are traversed by filaments filled with electrons, which are highlighted in blue.
In red, we see that the halos in the simulation lead to very high DM detections.
These halos dominate the shape of the high DM tail of $p(\mathrm{DM}|z)$.
By zooming in on single halos along one of the filaments using the pencil beam catalog, the effect of the circumgalactic medium on the DM signal becomes apparent.

In Figure \ref{fig:fullsky_evolution} we show the evolution of the full-sky map presented in Figure \ref{fig:fullskymap} as a function of redshift.
The more we evolve the map in redshift, the more DM is accumulated. 
The local environment visible at redshift $0.01$ is superimposed by signals from higher redshifts.
Consequently, the correlations visible at small modes at low redshifts are overlaid by larger modes the further away we go from the observer’s position.

The location of the observer was chosen as a Milky Way-like environment.
In Figure \ref{fig:fullskymap} on the right side of the HEALPix map we can see in red several very high DM clumps.
These clumps correspond to a Virgo-like cluster around $17$ Mpc away from the observer.
Close to the center of the map, we identify a more diffuse structure that has a big extend on the sky.
This halo corresponds to an Andromeda-like galaxy with a stellar mass of more than $3\times10^{10}\,M_{\odot}$ and at a distance of around $2000$ kpc from the observer.
For comparison, we show in Figure \ref{fig:fullsky_observer_change} five different Milky Way-like DM environments at redshift $0.01$.
All of which include a Virgo-like cluster and Andromeda-like galaxy in their surroundings.
For more details on how we identify the Milky Way-like environments, see Section \ref{subsec:observer_position}. 

In addition, we convolve our full-sky catalog with the footprints of the latest observational surveys to create simulated DM FRB observations on the full sky. 
To this end, we again use the sample of FRBs from \cite{Connor_2024}.
We compare our mock FRBs with the observational sample in Figure \ref{fig:mollview_real_mock_data}.

\begin{figure*}[]
    \centering
    \includegraphics[width=0.95\textwidth]{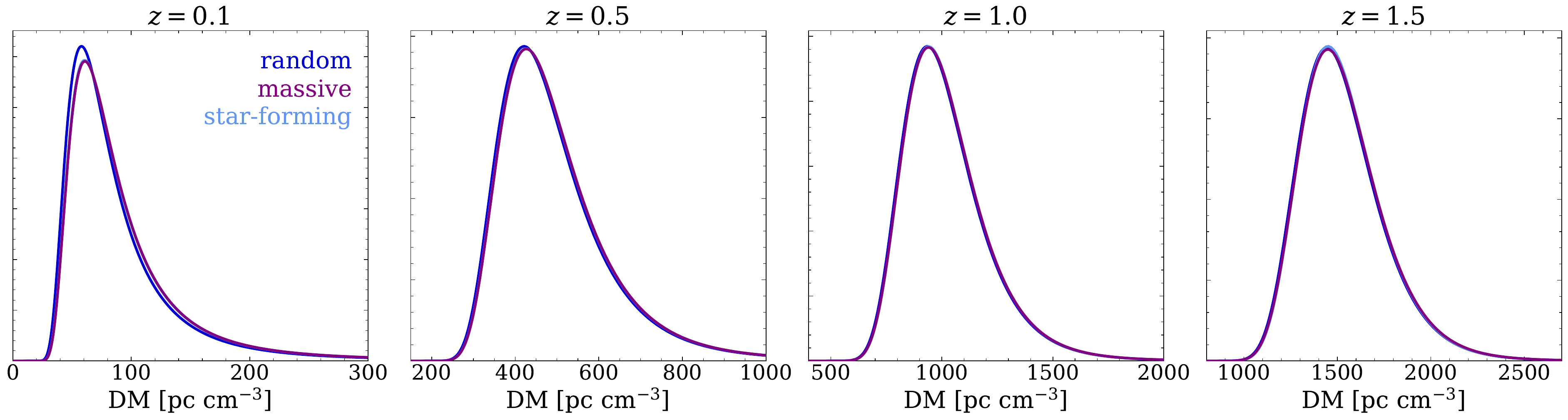}
    \caption{\textbf{DM distribution evolution including the effect of the FRB environment.}
    We show the effect of the FRB environment on our DM distribution at four different redshifts.
    We distinguish between two cases per redshift, i.e. we study FRBs from halos with a statistically high stellar mass (massive, purple) as well as halos with a statistically high star formation rate (star-forming, light blue).
    We see that the effect of the FRB environment or $\mathrm{DM_{\mathrm{cos,\,env}}}(z)$ is only significant for small redshifts ($z\approx0.1$).
    At higher redshifts the effect becomes negligible.
    \label{fig:effect_halos}}
\end{figure*}

Figure \ref{fig:mollview_real_mock_data} illustrates that our DM maps provide the necessary tools to not only statistically forecast FRB DMs but also to incorporate their spatial location in the sky.
This is especially important for a theoretical investigation of the FRB DM-galaxy angular cross-power spectrum.
Our maps are perfectly suited for this purpose thanks to their full-sky coverage, high number of pixels, and great integration depth, derived from a complete reconstruction of all traversed line segments within the Voronoi mesh.
Previous studies about the angular cross-power spectrum in IllustrisTNG \citep{Cheng_2025} have been limited by a coarse ray-tracing resolution and shallow integration depth (see also Figure \ref{fig:cartesian_gridding}).

Furthermore, Figure \ref{fig:mollview_real_mock_data} suggests that a full 3D reconstruction of the cosmic baryon web in some patches of the sky might be feasible with a spatially dense enough sample of local FRBs.
We will leave the task of performing such reconstructions using observed FRBs to future work, pointing out that our maps provide the necessary dataset to test reconstructions theoretically.

\subsection{Effect of the FRB Host Environment}\label{subsec:FRBhost}
While in this work we do not investigate the host or Milky Way contributions to the total DM signal, but treat those components with existing models, we can ask whether the extragalactic DM is biased by the location of the FRB in addition to its host or Milky Way contribution.
In all previous catalogs, we have located the FRBs at random positions in the simulation.
The FRBs come from all parts of the cosmic web, including halos, filaments, and voids.
Here, we investigate whether, in addition to the host contribution (that is not modeled in this work), there is a bias based on the location of the FRB in the large-scale structure.

For this purpose, we first select two sets of halos from which the FRBs originate, star-forming halos and massive halos (see Section \ref{subsec:startpoints_halo} for details).
For both halo types, we position mock FRBs at redshifts $0.1$, $0.5$, $1.0$, and $1.5$.
Next, we exclude the host contributions of the halos.
We define the host as all matter within three virial radii of the center of the corresponding halo.
Consequently, the signal we are left with contains only the additional extragalactic bias.

We show our results in form of histograms in Figure \ref{fig:effect_halos}.
We see that only at redshift $0.1$ there is a bias above the percent level.
The histograms become indistinguishable as soon as we investigate higher redshifts.
At these distances, the effect drops to the percent and sub-percent levels.
We conclude that the bias becomes negligible for redshifts larger than $0.1$.

Moreover, we find that, overall, the massive hosts exhibit a greater bias than the star-forming ones.
This can be explained by the fact that massive halos are likely to be surrounded by dense filaments, which are imprinted in the DM signal. 
In contrast, hosts selected according to their star formation rate are not necessarily located in the densest parts of the cosmic web.

\begin{figure*}[]
    \centering
    \includegraphics[width=1.0\textwidth]{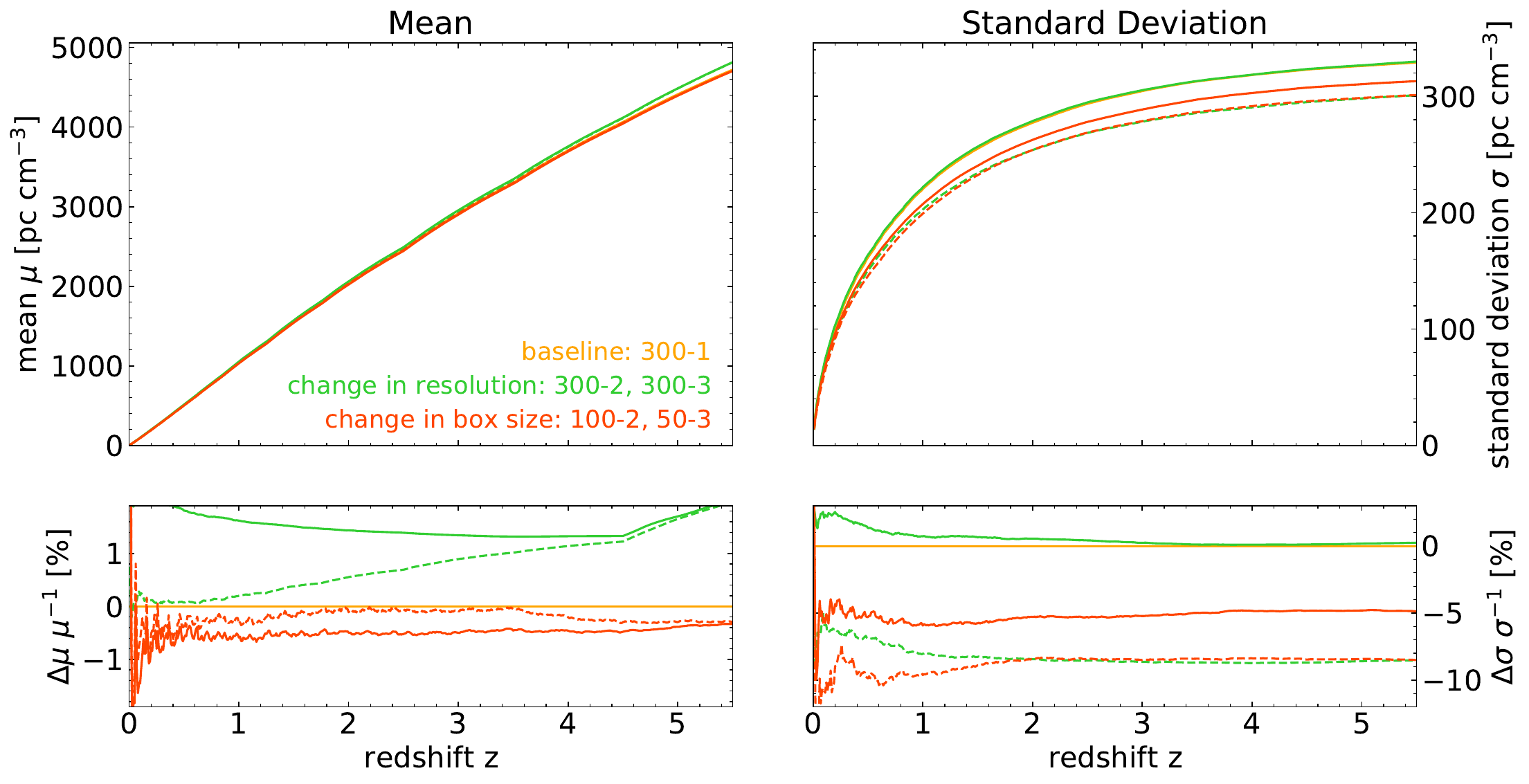}
    \caption{\textbf{Effect of box size and resolution.}
    We show the effect of changing the simulation box size and resolu tion.
    We use the stacked catalog for this analysis to avoid problems with a too short repetition length (see Section \ref{subsec:choicerayvector}).
    The baseline corresponding to TNG300-1 is shown in orange.
    We illustrate in orange-red the effect of changing the box size from $205 \,h^{-1}\,$Mpc to $75 \,h^{-1}\,$Mpc (solid) and to $35 \,h^{-1}\,$Mpc (dashed), while we keep the resolution roughly constant.
    This corresponds to using TNG100-2 and TNG50-3.
    In green we show the effect of changing the resolution.
    Using TNG300-2, which is based on $1250^3$ gas cells, is displayed by the solid line. Using TNG300-3, which is based on $625^3$ gas cells, is represented by the dashed line.
    \textbf{Left panel.}
    In terms of the mean of $p(\mathrm{DM}|z)$, both the resolution and size of the simulation, leave the distribution roughly unaffected.
    \textbf{Right panel.}
    In terms of the standard deviation, switching to TNG300-2 leaves $p(\mathrm{DM}|z)$ unchanged, whereas using TNG300-3 changes $\sigma$ by more than $8\%$. 
    The size of the simulation box also biases the standard deviation.
    The bias can be more than $8\%$ when comparing a box length of $205\,h^{-1}\,$Mpc with $35\,h^{-1}\,$Mpc.
    \label{fig:box_size_resolution}}
\end{figure*}

\subsection{Effect of the Simulation Box Size and Resolution}\label{subsec:size_resoluttion}
In this section we will analyze the effect of changing the simulation box size and resolution.
As outlined in Section \ref{subsec:choicerayvector}, when using box sizes smaller than the largest TNG simulation suite, TNG300, with $L = 205 \,h^{-1}\,$Mpc, the repetition length gets too short such that distortion effects will be imprinted into a continuous catalog as in the case when using the periodic method.
Therefore, the following discussion is based on the stacked method.

We change the box size from $205 \,h^{-1}\,$Mpc (TNG300-1) to $75 \,h^{-1}\,$Mpc (TNG100-2) and to $35 \,h^{-1}\,$Mpc (TNG50-3), where we choose TNG runs that are roughly equal in terms of their resolution.
In Figure \ref{fig:box_size_resolution} in orange-red we see with a solid (TNG100-2) and with a dashed line (TNG50-3) that the overall mean of the DM distribution is unaffected (varying by less than $1\%$) by a box size change.
However, the standard deviation changes by about $5\%$ and $8\%$ for TNG100-2 and TNG50-3, respectively, implying that the usage of smaller box sizes distorts the DM distribution. 
We also caution that we expect using runs with smaller box sizes, as for example implemented in the CAMELS project with $25\,h^{-1}\,$Mpc \citep{Medlock_2024, Sharma_2025}, to lead to a greater distortion of the DM distribution.

While we assumed throughout this paper that TNG300's volume is big enough for getting accurate DM result, the deviation in the variance suggests that larger volumes ($L > 205\,h^{-1}\,$Mpc) will have a small additional impact.
Future studies, relying on even bigger volumes like MillenniumTNG \citep{MTNG_Cesar_2023, MTNG_Pakmor_2023, MTNG_Barrera_2023, Ferlito_2023_MTNG, Delgado_2023_MTNG}, will be able to determine this effect. 

In green, we show what happens if we keep the volume the same but change the resolution.
To this end we use the TNG300-2 and TNG300-3 runs that both have a side length of $205 \,h^{-1}\,$Mpc but decrease in resolution.
We see that the mean deviates by at most $2\%$ in both the medium (TNG300-2, green solid line) and low resolution (TNG300-3, green dashed line) cases.
In the case of TNG300-2 (resolution: $6\times 10^7\,h^{-1}\,M_{\odot}$) also the variance is almost unaffected compared to TNG300-1 (resolution: $8\times 10^6\,h^{-1}\,M_{\odot}$).
In contrast, the lowest resolved run we analyse, TNG300-3 (resolution: $5\times 10^8\,h^{-1}\,M_{\odot}$), shows a change in the standard deviation of the DM distribution by $8\%$.
The effect is similar to the distortion introduced by using a box size of $35 \,h^{-1}\,$Mpc \mbox{(TNG50-3)}.

We conclude that, for most applications, TNG300-2 is consistent with TNG300-1.
However, in cases of a very low resolution with baryonic masses of less than $5\times 10^8\,h^{-1}\,M_{\odot}$ or a box size smaller than $35\,h^{-1}\,$Mpc, there is a distortion of the distribution that may affect results drawn from such runs.

\section{Summary and Outlook}\label{sec:summary_outlook}
In this work, we created over $20$ different dispersion measure (DM) catalogs using the IllustrisTNG simulation suite to examine the cosmological evolution of Fast Radio Burst (FRB) signals.
Our main findings are as follows:

\begin{enumerate}

    \item We present four different methods to extract FRB DMs from simulations which we refer to as periodic, trapezoidal, stacked, and continuous.
    The latter two, stacked and continuous, are the only ones leading to accurate results.
    We recommend the continuous method for large volume simulations and the stacking approach for small volume simulations.

    \item We study the shape of $p(\mathrm{DM}|z)$ and find that a modified-normal density function with a power law tail, described by equation (\ref{eq:pdf_np}), with shape parameters $\alpha=1$ and $\beta=3.3$ provides a very good fit to the distribution across all redshifts from $0$ to $5.5$.
    Previously proposed shape parameters of $\alpha = 3$ and $\beta=3$ or a log-normal distribution are not favored by the data. 

    \item We analyze $p(\mathrm{DM}|z)$ in light of the DM-redshift relation and compare our continuous ray-tracing results to a recent observational FRB sample from the DSA-110, ASKAP, and CHIME.
    We find our results to be consistent with the observational data.
    
    \item We also confirm our ray-tracing results with an analytical model and commented on the effect of feedback and the evolution of the UV background using the fraction function between dispersive electrons and baryons $f_{eb}(z)$.
    
    \item We create full-sky DM maps centered on Milky Way-like environments.
    Using these full-sky maps, we generate mock signals, incorporating the observed FRB locations in the sky, and compare them to observational data.

    \item We show that, in addition to the pure host contributions, large-scale correlations between the host location and the cosmic web can bias FRB signals at small redshifts ($z\approx0.1$).
    FRBs originating from halos selected by their stellar mass show generally a larger effect than halos selected by their star formation rate.
    This effect can be neglected at redshifts larger than $0.1$.
  
    \item We find that the simulation box size and resolution can affect the DM distribution.
    Small box sizes of $75\,h^{-1}\,$Mpc or $35\,h^{-1}\,$Mpc can distort the results by $5\%$ and $8\%$, respectively. 
    Decreasing the resolution from $8\times 10^6\,h^{-1}\,M_{\odot}$ to $6\times 10^7\,h^{-1}\,M_{\odot}$ does not significantly affect the FRB DM distribution.
    However, lower resolutions with baryonic masses of less than $5\times 10^8\,h^{-1}\,M_{\odot}$ can distort the distribution by more than $8\%$. 
    We expect these effects to increase with smaller boxes and lower resolutions.

\end{enumerate}

Looking to the future, it will be essential to study the effects of different feedback implementations and the evolution of the UV background in interplay with the overall underlying cosmology.
One way to address these questions is to examine the same simulation setup under different initial conditions, as well as other models, such as EAGLE \citep{Eagle_2015_Crain, Eagle_2015_Schaye} and SIMBA \citep{Simba_2019}.
A suite of simulations like CAMELS \citep{CAMELS_presentation} is a promising tool for taking steps in this direction.

Furthermore, we have now entered the era of observing correlations between FRB DMs and the underlying galaxy distribution.
This can be achieved through stacking analyses, which combine DM and galaxy information, as discussed in \citep{Hussaini_2025_corr}.
Alternatively, one can take an approach to measure the FRB DM-galaxy angular cross-power spectrum \citep{Wang_2025_CHIME_Xcorr}.
In both cases, our full-sky catalog is the ideal theoretical counterpart for making forecasts and investigating potential biases or systematic uncertainties.
Based on our catalogs, future work should be able to determine whether obtaining a large number of non-localized FRBs from instruments like CHIME or a smaller number of precisely localized FRBs from a telescope like the DSA-110 is more advantageous in the short term.

\section*{Data and Code Availability}\label{sec:data_access}

We make the DM catalogs derived in this work publicly available and easily accessible to the community. The catalogs can be downloaded at \href{https://ralfkonietzka.github.io/fast-radio-bursts/ray-tracing-catalogs/}{https://ralfkonietzka.github.io/fast-radio-bursts/ray-tracing-catalogs/}.

The code used to derive the results presented in this work is available upon reasonable request.

\section*{Acknowledgements}
We would like to thank Daniel Eisenstein, Charlie Conroy, Shy Genel, Adam J. Batten, Michal Jaroszynski, Maryam Hussaini, Ryuichi Takahashi, Ruediger Pakmor, Oliver Zier, Jake Bennett, Scott Lucchini, Claire Lamman, Carolina Cuesta-L\'azaro, Tanveer Karim, Simon White, Vikram Ravi, Kritti Sharma, Shion E. Andrew, Kiyoshi W. Masui, Haochen Wang, Robert Reischke, Calvin Leung, Aaron Smith, Vedant Chandra, MohammadReza Torkamani and Sarah Jeffreson for helpful correspondence and discussions.
The analysis presented in this paper was performed on the FASRC Cannon cluster supported by the FAS Division of Science Research Computing Group at Harvard University.
V.S. and L.H. acknowledge support by the Simons Collaboration on ``Learning the Universe''.
Support for V.S. was provided by Harvard University through the Institute for Theory and Computation Fellowship.
The analysis presented in this paper was aided by the following software packages: \textsc{numpy} \citep{numpy_ndarray, numpy_Harris_2020}, \textsc{scipy} \citep{scipy_Virtanen_2020}, \textsc{matplotlib} \citep{Hunter_matplotlib}, \textsc{numba} \citep{Lam_Numba}, \textsc{healpy} \citep{Gorski_2005_healpy, Zonca2019_healpy}, \textsc{astropy} \citep{Astropy_2013, Astropy_2018, Astropy_2022}, \textsc{tqdm} (\href{https://github.com/}{https://tqdm.github.io/}), \textsc{h5py} (\href{http://www.h5py.org/}{http://www.h5py.org/}), and \textsc{plot.ly} (\href{https://plotly.com}{https://plotly.com}).
We have also used the Astrophysics Data Service (\href{http://adsabs.harvard.edu/abstract_service.html}{ADS}) and the \href{https://arxiv.org}{arXiv} preprint repository during this project and writing of the paper.

\bibliography{bibliography}{}
\bibliographystyle{aasjournal}



\end{document}